\documentclass[reprint,aps,prb,floatfix,longbibliography]{revtex4-1}
\usepackage{color,graphicx,calc,url}
\usepackage{xcolor}
\usepackage{dcolumn}
\usepackage{bm}
\usepackage{amssymb}
\usepackage{amsmath}
\usepackage{wasysym}
\usepackage{mathrsfs} 
\usepackage{array}
\usepackage{mathtools}
\usepackage{commath}
\usepackage{adjustbox}
\usepackage{braket}
\usepackage{float}
\usepackage{textcomp}
\usepackage{makecell}

\usepackage[colorlinks=true,
            linkcolor=red,
            urlcolor=blue,
            citecolor=blue]{hyperref}

\begin{document}
\title{Cubic-in-magnetization contributions to the magneto-optic Kerr effect investigated for Ni(001) and Ni(111) thin films}

\author{Robin Silber$^{1,2,*,}$\textsuperscript{\textdagger}, Maik Gaerner$^{3,}$\textsuperscript{\textdagger}, Kamil Postava$^{1}$, Jaroslav Hamrle$^{4,5}$, Timo~Kuschel$^{3,6}$ }
\affiliation{$^1$Department of Materials Engineering and Recycling, Faculty of Materials Science and Technology, VSB-Technical University of Ostrava, Ostrava 70800, Czech Republic\\
$^2$ Nanotechnology Centre, Centre for Energy and Environmental Technologies, VSB-Technical University of Ostrava, Ostrava 70800, Czech Republic\\
$^3$ Faculty of Physics, Bielefeld University, Universit\"atsstra\ss e 25, 33615 Bielefeld, Germany\\
$^4$ Faculty of Mathematics and Physics, Charles University, Ke Karlovu 5, 12116 Prague, Czech Republic\\
$^5$ Faculty of Nuclear Sciences and Physical Engineering, Czech Technical University, Trojanova 13, 12000 Prague, Czech Republic\\
$^6$ Institute of Physics, Johannes Gutenberg University Mainz, Staudingerweg 7, 55128 Mainz, Germany}

\date{\today}
\email{Contact author: robin.silber@vsb.cz}
\thanks{\textsuperscript{\textdagger}These authors contributed equally to this work.}


\begin{abstract}
The existence of higher-order effects in magnetization in magneto-transport and magneto-optic phenomena is well known. These effects are of importance not only in research but also directly in applications. Anisotropic magneto-resistance, anisotropic magneto-thermopower, quadratic magneto-optic Kerr effect (QMOKE), Voigt effect or x-ray linear magnetic dichroism (and birefringence) - all of these are effects of second order in magnetization. Recently, we have reported on first systematic observations of the cubic-in-magnetization magneto-optic Kerr effect (CMOKE) in Ni(111) thin films. In this paper, we introduce the detailed theory of CMOKE by deriving the magneto-optic tensor of third order in magnetization, denoted as $\bm{H}$, and comparing the strength of CMOKE for different crystal orientations theoretically and experimentally. In crystals with cubic symmetry, the tensor $\bm{H}$ is described by two independent parameters $H_{123}$ and $H_{125}$. Together with the linear magneto-optic tensor $\bm{K}$ and quadratic magento-optic tensor $\bm{G}$, the permittivity tensor is described up to third order in magnetization. We analytically describe equations of the magneto-optic Kerr effect (MOKE) including the contribution of QMOKE and CMOKE itself for (001)- and (111)-oriented cubic crystal structures. Those are compared to experimental measurements of two samples with an (001)- and (111)-oriented fcc Ni layer, respectively. The experimental data are measured using the so-called eight-directional method, which was developed to separate linear and quadratic MOKE contributions and to observe their individual anisotropic dependence on the crystal lattice direction, but is applicable to CMOKE contributions as well. Further, we use Yeh's 4$\times$4 transfer matrix calculus to simulate and describe the experimental measurements phenomenologically from the permittivity tensor up to third order in $\bm{M}$. This allows us to obtain the values of the magneto-optic parameters of the Ni layer for both crystal orientations. We find that the MOKE anisotropy that stems from the magneto-optic tensor $\bm{H}$ described as $\Delta H = H_{123}-3H_{125}$, is much more pronounced for the (111)-oriented cubic crystal structure, for which it manifests as three-fold in-plane angular dependencies of MOKE with longitudinal and also with transversal magnetization direction, respectively. For (001)-oriented cubic crystal structures, $\Delta H$ should manifest as a four-fold angular dependence of the MOKE with longitudinal and transversal magnetization directions, but its amplitude is predicted to be smaller than in the case of the (111)-oriented cubic crystal structure, which may explain why the CMOKE has not yet been experimentally identified in (001)-oriented cubic crystal structures.

\end{abstract}

\maketitle
\section{Introduction}

The magneto-optic Kerr effect (MOKE) is one of the most established tools in magnetic material research \cite{J.Kerr1877, MOOG1985, Qiu2000, Buschow2001}. It can be used to probe the magnetization of ferro-, ferri- and antiferromagnetic  materials \cite{Krinchik68, Visnovsky1995, Hamrle2002, Kuschel2011, Kuschel2012, Bergman2006, Saidl2017a, Nemec2018} in applications such as measurements of the magnetic anisotropy \cite{NEGRE1999, Mattheis99}, vectorial MOKE measurements \cite{Florczak1990, Vavassori2000, Kuschel2011}, MOKE spectroscopy \cite{Krinchik68, Ferguson69, Yaresko98, Visnovsky1999, Silber18, Silber19, Silber20} or Kerr microscopy \cite{Schafer07, McCord15, Janda18, Kim20}.  With the development of ultrashort pulsed lasers, MOKE has been employed in time-resolved pump-probe techniques as well \cite{Beaurepaire1996, Tesarova12, Janda18, Zhao21}. Here, magnetization dynamics can be observed with time resolution down to fs. One of the recent cutting-edge applications of MOKE is the measurement of spin-orbit torque (SOT) effective fields \cite{Garello13, Fan14, Montazeri15, Stamm17} and the observation of magnetization dynamics and magnetization switching in spintronic terahertz emitters by SOTs generated by ps electric pulses \cite{Jhuria20}. The latter application connects spintronics and terahertz research, both being in the tight focus of the last decade of research. 

In a lot of cases mentioned above, only the linear dependence of MOKE on the magnetization $\bm{M}$ is assumed. Although the broad community is usually well aware of the existence of higher-order effects of MOKE, they are rarely considered in the model. In the past, systematic investigations of quadratic MOKE (QMOKE), being proportional to $\bm{M}^2$, have been published \cite{Visnovsky1986, Carey98, Mertins2001, Hamrlova2013}, mostly involving  methods for the separation of linear MOKE (LinMOKE) and QMOKE \cite{Mattheis99, Postava2002, Mewes2004, Liang2015, Liang2016}, methods for investigating Heusler compounds using QMOKE \cite{Hamrle07a, Hamrle07b, Gaier2008, Muduli08, Muduli09, Trudel10c, Trudel11,  Wolf2011, Lobov2012, Silber20}, enhanced vectorial magnetometry \cite{Kuschel2011,  Kuschel2012, Tesarova12, Fan16}, QMOKE spectroscopy techniques \cite{Sepulveda2003, Hamrlova2016, Silber18, Silber19} and others \cite{Buchmeier2009, Valencia2010}. Magnetic linear birefringence and dichroism (MLB and MLB, respectively) during reflection \cite{Metzger65} are a case of QMOKE appearing for a normal angle of incidence (AoI), where the set of polarization eigenmodes consists of linear polarizations that are parallel and perpendicular to the $\bm{M}$. Note, that if the MLB and MLD stems from gradient of magnetization $\nabla \bm{M}$, the underlying effect is known as Sch\"{a}fer-Hubert effect \cite{Schafer90, Schafer95, Krambersky92}. Often, QMOKE is a rather unwanted contribution to their MOKE measurements, which is oftentimes filtered out, as the linear dependence between MOKE and $\bm{M}$ simplifies experimental analysis. Nevertheless, there are actual applications for which the properties of QMOKE itself and other magneto-optic effects of second order in magnetization are leveraged. One example is QMOKE-based magneto-optic microscopy \cite{McCord15, Janda18}, or the above mentioned measurements of SOT effective fields and time-resolved magnetization dynamics \cite{Montazeri15}, as well as the observation of optical SOT \cite{Tesarova13}, for which information about the in-plane component of the magnetization is carried by the QMOKE signal. 

Further, QMOKE measurements of antiferromagnetic (AFM) materials have been realized \cite{Saidl2017a, Nemec2018}, for which LinMOKE measurements cannot be used due to a lack of net magnetization in most AFMs. As the control of AFM spin orientation by SOTs has been demonstrated \cite{Jungwirth2016, Wadley2016}, there will be increasing demand for fast and easily accessible methods for characterization of the N\'{e}el vector in AFMs. Hence, the motivation for systematic investigation of higher-order MOKE effects is clearly in place, and may bring enhanced capabilities of MOKE, either by acknowledging a higher-order model instead of the simple linear dependence of the  MOKE on $\bm{M}$, or by the new utilities that the MOKE of higher-order can provide and that cannot be yielded by LinMOKE itself.  

In addition to QMOKE, cubic MOKE (CMOKE) being the contribution to MOKE of third order in $\bm{M}$, has promising potential applications but its contribution has not been considered during MOKE data processing so far. In our previous work \cite{Gaerner24}, we showed the angular dependence of CMOKE for (111)-oriented cubic crystal structure and discussed how structural domain twinning in (111)-oriented Ni thin films affects the amplitude of the CMOKE contribution. From our point of view, the CMOKE contribution to the overall MOKE signal is important because \textit{(i)} it has a finite angular dependence in contrast to LinMOKE and it is superimposed to the longitudinal MOKE (LMOKE) contribution. Thus it can affect the evaluation of LMOKE data, if not  taken into account and not treated carefully. Note that LMOKE (odd in $\bm{M}$) and CMOKE (odd in $\bm{M}$) cannot be separated as easily as LMOKE (odd in $\bm{M}$) and QMOKE (even in $\bm{M}$). \textit{(ii)} It could be used to tailor applications such as CMOKE microscopy for correlating magnetic domains with structural domain twinning, or time evolution of twinning by time resolved MOKE, both based on CMOKE angular dependence amplitude detection. \textit{(iii)} For (111)-oriented cubic samples, CMOKE has a finite value at normal AoI, whereas LMOKE does not \cite{Gaerner24, pan25}. This can be leveraged for vectorial MOKE measurements at normal AoI to obtain information about the in-plane component of $\bm{M}$, while the polar MOKE (PMOKE) would provide information about the out-of-plane component of $\bm{M}$. Although for the latter application, QMOKE is being used already \cite{Montazeri15, Tesarova12, Tesarova13, Fan16}, the advantage of CMOKE over QMOKE is that CMOKE is odd in magnetization, thus providing information about the direction of magnetization in the plane as well. Also, it may happen, that at particular material and wavelength, CMOKE is dominant contribution over QMOKE. Therefore, we believe that the CMOKE is worth to be systematically investigated in further detail. 

In this paper, we provide \textit{(i)} expressions of the permittivity tensor described up to third order in $\bm{M}$ for both (001)-oriented cubic crystal structures and (111)-oriented cubic crystal structures at a general sample in-plane orientation angle $\alpha$, \textit{(ii)} relations describing the eight-directional method for both (001)-and (111)-oriented cubic crystal structures extended up to third order in $\bm{M}$ \textit{(iii)} experimental measurements via the eight-directional method for an MgO(001)//Ni/Pt sample, denoted as sample Ni(001) in the text, and, experimental measurements of the eight-directional method for an MgO(111)//Ni/Si/SiO$_x$ sample, denoted as sample Ni(111) in the text. \textit{(iv)} The experimental data are described by two models - an analytical model developed in Sec.~\ref{theory_moke} and by Yeh's 4$\times$4 transfer matrix formalism \cite{Yeh1980}, describing the propagation of light in a multilayer structure consisting of generally optically anisotropic layers. The latter model allows to obtain values of the magneto-optic parameters, that describe the change of the permittivity tensor up to third order in $\bm{M}$. The results from both samples are compared and thoroughly discussed in Sec. \ref{experimental observations}.

%%%%%%%%%%%%%%%%%%%%%%%%%%%%%%%%%%%%%%%%%%%
%-----THEORY OF LINEAR AND QUADRATIC MOKE
%%%%%%%%%%%%%%%%%%%%%%%%%%%%%%%%%%%%%%%%%%%
\section{Theory of linear, quadratic and cubic MOKE}
\label{theory_moke}
\subsection{Introduction to the permittivity of a magnetized crystal}
The complex Kerr angle $\Phi_{s/p}$ for $s$ and $p$ polarized incident light is described  as \cite{Visnovsky06, Hecht2002}

\begin{subequations}
\begin{align}	
\Phi_{s}&{}=\theta_{s}+i\epsilon_{s}\approx
\frac{\tan{\theta_{s}}+i\tan{\epsilon_{s}}}{1-i\tan{\theta_{s}}\tan{\epsilon_{s}}}=-\frac{r_{ps}}{r_{ss}},
\\[5mm]
\Phi_{p}&{}=\theta_{p}+i\epsilon_{p}\approx\frac{\tan{\theta_{p}}+i\tan{\epsilon_{p}}}{1-i\tan{\theta_{p}}\tan{\epsilon_{p}}}=
\frac{r_{sp}}{r_{pp}}\quad.
\end{align}
\label{Kerr_basic}
\end{subequations}

\noindent
Here, $\theta_{s/p}$ and $\epsilon_{s/p}$ are the Kerr rotation and Kerr ellipticity, respectively. As these Kerr angles are generally very small  ($<1^{\circ}$), the small angle approximation \cite{Buschow2001} is used in Eq.~(\ref{Kerr_basic}).  

The reflection coefficients $r_{ss}, r_{ps}, r_{sp}, r_{pp}$ in Eq.~(\ref{Kerr_basic}) do fundamentally depend on the permittivity tensor $\bm{\varepsilon}$. We can relate the complex Kerr angles to the permittivity of a thin ferromagnetic layer with good approximation as \cite{Hamrle03}
\begin{subequations}
\begin{align}
\Phi_s &{}=-\frac{r_{ps}}{r_{ss}}= A_{s}\left(\varepsilon_{yx}-\frac{\varepsilon_{yz}\varepsilon_{zx}}{\varepsilon_d}\right)+B_s\varepsilon_{zx},\\[5mm]
\Phi_p &{}=\frac{r_{sp}}{r_{pp}}=-A_{p}\left(\varepsilon_{xy}-\frac{\varepsilon_{zy}\varepsilon_{xz}}{\varepsilon_d}\right)+B_p\varepsilon_{xz},
\end{align}
\label{Kerr_analyt}
\end{subequations}

\noindent
with $A_{s/p}$ and $B_{s/p}$ as the optical weighting factors, which include the effect of the thickness of the magnetic layer, the influence of the other layers in the stack, and the dependence on the AoI. $A_{s/p}$  is an even function of the AoI, and persists with normal AoI, while $B_{s/p}$ is an odd function of the AoI and vanishes with normal AoI. The change of the permittivity tensor with magnetization is then the phenomenological origin of MOKE \cite{Hecht2002, Visnovsky06}. We can describe the permittivity tensor of magnetized ferromagnetic medium as the Taylor series $\bm{\varepsilon}=\bm{\varepsilon}^{(0)}+\bm{\varepsilon}^{(1)}+\bm{\varepsilon}^{(2)}+\bm{\varepsilon}^{(3)}+... $, where the superscript denotes the order in $\bm{M}$.  To successfully explain all the experimental observations in this work, we developed the permittivity tensor up to third order in magnetization as 

\begin{equation}
	\varepsilon_{ij}=\varepsilon_{ij}^{(0)}+\underbrace{K_{ijk}M_k}_{\varepsilon_{ij}^{(1)}}+\underbrace{G_{ijkl}M_k M_l}_{\varepsilon_{ij}^{(2)}}+\underbrace{H_{ijklm}M_k M_l M_m}_{\varepsilon_{ij}^{(3)}}
\label{eq:permittivity_full}
\end{equation}

\noindent
with $M_k$, $M_l$ and $M_m$ being the components of the normalized magnetization
\begin{equation}
\bm{M}=
\begin{bmatrix}
M_1\\
M_2\\
M_3	
\end{bmatrix}
=
\begin{bmatrix}
M_x\\
M_y\\
M_z	
\end{bmatrix}
=
\begin{bmatrix}
M_T\\
M_L\\
M_P	
\end{bmatrix}.
\label{eq:mag_vec}
\end{equation}

\noindent
Here, the subscripts $x,y,z$ coincide with the directions in our $\hat{x}$, $\hat{y}$, $\hat{z}$  coordinate system  and the subscripts $T,L,P$ refer to three basic MO configurations, being transversal, longitudinal and polar, respectively \cite{Visnovsky06} (see Appendix~\ref{app:conventions} with its Fig. \ref{f:xyz_sys}). Elements $\varepsilon_{ij}^{(0)}$  present the non-perturbed permittivity tensor of a non-magnetized medium, whereas  $K_{ijk}$ and $G_{ijkl}$ are the elements of the so-called linear and quadratic MO tensors $\bm{K}$ and $\bm{G}$ of third and fourth rank, which creates contributions $\varepsilon^{(1)}$ and $\varepsilon^{(2)}$ to the overall permittivity, respectively \cite{Visnovsky1986}. For a cubic crystal structure with inversion symmetry, the elements of those tensors are simplified as \cite{Visnovsky1986}

\begin{subequations}
\begin{align}
 	\varepsilon_{ij}^{(0)} &{}= \,\delta_{ij}\varepsilon _{d},\label{eq:perm_upto2_a}\\[1mm]
 	K_{ijk} &{}=\, \epsilon_{ijk}K, \\[1mm]
 	G_{iiii} &{}=\, G_{11}, \\[1mm]
 	G_{iijj} &{}=\, G_{12}, \qquad i\neq j,\\[1mm]
 	G_{1212} &{}=\, G_{1313}=G_{2323}=G_{44},\label{eq:perm_upto2_e}
\end{align}
\end{subequations} 

\noindent
with  $\delta_{ij}$ and $\epsilon_{ijk}$ being the Kronecker delta and the Levi-Civita symbol, respectively. Thus, for cubic crystal structures, ${\varepsilon}^{(0)}_{ij}$ is described by a scalar $\varepsilon_d$. The linear MO tensor $\bm{K}$ is isotropic (i.e. is rotationally invariant) and is described by one independent parameter $K$. For the quadratic MO tensor $\bm{G}$ there are two independent parameters in the case of cubic crystal structures, being $G_s=(G_{11}-G_{12})$ and $2G_{44}$. As $G_{11}$ and $G_{12}$ emerge only with the $M_i^2$ element of $\bm{M}$, the experiment can determine solely the difference $G_{11}-G_{12}$. To separate those phenomenological parameters, one would need to change the length of the magnetization vector. The parameter $\Delta G=G_s-2G_{44}$ denotes the anisotropic strength of this $\bm{G}$ tensor  \cite{Hamrle07b, Kuschel2011}. The abbreviated subscripts 11, 12, and 44 have been chosen from the individual position of elements in the Voigt notation of the $\bm{G}$ tensor.

Finally, $H_{ijklm}$ are the components of the MO tensor $\bm{H}$ of the fifth rank, being cubic in the magnetization. Derivation steps of the $\bm{H}$ tensor are discussed in Appendix \ref{app:H_tensor}. The third-order MO tensor $\bm{H}$ possesses two independent parameters, which are $H_{123}$ and $H_{125}$. The abbreviated subscripts 123 and 125 will be explained later on. Tensor $\bm{H}$ is then constructed as 
\begin{subequations}
\begin{align}
	&H_{ijkkk}=\epsilon_{ijk}H_{123},\label{eq:H_tensor_short_a}\\[1mm]
	&H_{ijkll}=\epsilon_{ijk}H_{125}, \\[1mm]
	&H_{ijklm}=0, \qquad k \neq l \neq m .\label{eq:H_tensor_short_c}
\end{align}
\end{subequations} 

\noindent
We further define an anisotropy parameter $\Delta H=H_{123}-3H_{125}$, which describes the anisotropy of the MO tensor $\bm{H}$. Our findings are in correspondence with Pethukov et al. \cite{Petukhov98}. The permittivity of the third order in magnetization $\varepsilon_{ij}^{(3)}$ is then constructed as the contraction of $H_{ijklm}M_kM_lM_m$, which is written in Voigt notation \cite{Gaerner24} as
\begin{widetext}
\begin{equation}
\begin{bmatrix}
\varepsilon_{23}^{(3)}\\
\varepsilon_{31}^{(3)}\\
\textcolor{red}{\varepsilon_{12}^{(3)}}\\	
\varepsilon_{32}^{(3)}\\
\varepsilon_{13}^{(3)}\\
\varepsilon_{21}^{(3)}
\end{bmatrix}
=
\begin{bmatrix}
H_{123}&0&0 & 0&0&3H_{125}&0&3H_{125}&0&0 \\
0&H_{123}&0 & 3H_{125}&0&0&0&0&3H_{125}&0 \\
0&0&H_{\textcolor{red}{12}\textcolor{blue}{3}} & 0&3H_{\textcolor{red}{12}\textcolor{green}{5}}&0&3H_{125}&0&0&0 \\
-H_{123}&0&0 & 0&0&-3H_{125}&0&-3H_{125}&0&0 \\
0&-H_{123}&0 & -3H_{125}&0&0&0&0&-3H_{125}&0 \\
0&0&-H_{123} & 0&-3H_{125}&0&-3H_{125}&0&0&0
\end{bmatrix}
\begin{bmatrix}
M_1^3\\
M_2^3\\
\textcolor{blue}{M_3^3}\\
M_2M_3^2\\
\textcolor{green}{M_3M_1^2}\\
M_1M_2^2\\
M_3M_2^2\\
M_1M_3^2\\
M_2M_1^2\\
M_1M_2M_3\\
\end{bmatrix}.
\label{eq:H_tensor_long}
\end{equation}
\end{widetext}

\noindent
Here, the indices 12 of $H_{123}$ and $H_{125}$ are indicating the first off-diagonal position in the permittivity tensor $\textcolor{red}{\varepsilon_{12}}$. Furthermore, the indices \textcolor{blue}{3}  and \textcolor{green}{5} are linked to the respective position in the $\bm{M}$ vector of the Voigt notation in Eq.~\eqref{eq:H_tensor_long}. Thus, the third order contribution of the first off-diagonal element of the permittivity tensor is $\textcolor{red}{\varepsilon_{12}^{(3)}}=H_{\textcolor{red}{12}\textcolor{blue}{3}}\textcolor{blue}{M_3^3}+3H_{\textcolor{red}{12}\textcolor{green}{5}}\textcolor{green}{M_3M_1^2}+...$ . Since there are only two independent parameters, all other non-zero entries of the $H$ tensor are equal to $H_{123}$ and $3H_{125}$.

Finally, note that all MO tensors discussed above are defined for the reference orientation of an (001)-oriented cubic crystal structure (see Appendix \ref{app:conventions} for the definition of this reference orientation). The rotation transformation of the MO tensors is then defined as
\begin{subequations}
\begin{align}
	G'_{ijkl}&= a_{in}a_{jo}a_{kp}a_{lq}G_{nopq},\label{eq:g_rot}\\	
	H'_{ijklm}&= a_{in}a_{jo}a_{kp}a_{lq}a_{mr}H_{nopqr}.
	\label{eq:h_rot}
\end{align}
\end{subequations}
with $a_{xy}$ being the elements of the orthogonal rotation matrix $\bm{a}$. The $G_{nopq}$ and $H_{nopqr}$ are elements of the tensors $\bm{G}$ and $\bm{H}$ before transformation, while $G'_{ijkl}$ and $H'_{ijklm}$ are elements of the tensors $\bm{G'}$ and $\bm{H'}$ after the transformation. Note that $\bm{\varepsilon}^{(0)}$ and $\bm{K}$ are isotropic for cubic crystal structures, thus they are invariant to such rotational transformations.

\subsection{Analytical equations of MOKE for (001)-oriented cubic crystal structures}

In this section, we derive the analytical expressions for the Kerr angles for (001)-oriented cubic crystal structures.  First, we will fully describe the off-diagonal elements of the permittivity tensor up to third order in $\bm{M}$, as shown in Eq.~\eqref{eq:permittivity_full}. To include the dependence on the in-plane orientation of the sample, the rotation about the $\hat{z}$-axis by a general angle $\alpha$ (see Eq.~\ref{eq:a_rot_z} of Appendix~\ref{app:conventions}) is applied to tensors $\bm{G}$ and $\bm{H}$ according to Eqs.~\eqref{eq:g_rot} and \eqref{eq:h_rot}. The off-diagonal elements of the permittivity tensor for (001)-oriented cubic crystal structures can then be evaluated as

\begin{equation}
\begin{split}
\varepsilon^{(001)}_{xy/yx}=\qquad \qquad \qquad \quad \pm K\quad &M_P\\
\left[2G_{44}+\frac{\Delta G}{2}(1-\cos(4\alpha))\right]\quad &M_LM_T\\
+ \frac{\Delta G}{4}\sin(4\alpha)\quad  &(M_T^2-M_L^2)\\
\pm H_{123}\quad &M_P^3 \\
\pm 3H_{125}\quad &M_PM_T^2\\
\pm 3H_{125}\quad &M_PM_L^2,
\end{split} 
\label{eq:eps_xy001}
\end{equation}

\begin{equation}
\begin{split}
\varepsilon^{(001)}_{zx/xz}= \qquad \qquad \qquad \quad \pm K \quad &M_L \\
+2G_{44}\quad &M_PM_T \\
\pm\frac{\Delta H\sin(4\alpha)}{4}\quad &M_T^3\\
\pm\frac{\Delta H\cos(4\alpha) +3H_{123}+3H_{125}}{4}\quad &M_L^3\\
\pm 3H_{125}\quad &M_LM_P^2\\
\mp\frac{3\Delta H\sin(4\alpha)}{4}\quad &M_TM_L^2\\
\mp\frac{3\Delta H\cos(4\alpha)-3H_{123}-3H_{125}  }{4}\quad &M_LM_T^2,
\end{split}
\label{eq:eps_xz001}
\end{equation}

and
\begin{equation}
\begin{split}
\varepsilon^{(001)}_{yz/zy}=\qquad \qquad \qquad \quad\pm K\quad &M_T\\
+2G_{44}\quad &M_LM_P \\
\pm\frac{\Delta H\cos(4\alpha)+3H_{123}+3H_{125}}{4}\quad &M_T^3\\
\mp\frac{\Delta H\sin(4\alpha)}{4}\quad &M_L^3\\
\mp\frac{3\Delta H\cos(4\alpha)-3H_{123}-3H_{125}}{4}\quad &M_TM_L^2\\
\pm 3H_{125}\quad &M_TM_P^2\\
\pm\frac{3\Delta H\sin(4\alpha)}{4}\quad &M_LM_T^2.
\end{split}
\label{eq:eps_yz001}
\end{equation}

\noindent

Using those permittivity elements in Eq.~\eqref{Kerr_analyt} provides the dependencies of the Kerr angles on \textit{(i)} MO parameters, \textit{(ii)} in-plane magnetization components and \textit{(iii)} sample orientation $\alpha$, for (001)-oriented thin films with cubic crystal structure. We derive
\begin{equation}
\begin{split}	
&\Phi_{s/p\, [M_P=0]}^{(001)} =\\
&\pm A_{s/p}\left[\left(2G_{44}-\frac{K^2}{\varepsilon_d}+\frac{\Delta G}{2}(1-\cos(4\alpha))\right) M_LM_T\right.\\
& \qquad \qquad \left.+\left(\frac{\Delta G}{4}\sin(4\alpha)\right) (M_T^2-M_L^2)\right]\\
&\pm B_{s/p}\left[K M_L+\frac{\Delta H\cos(4\alpha)+3H_{123}+3H_{125}}{4}M_L^3 \right.\\
&\qquad\qquad-\frac{3\Delta H\cos(4\alpha)-3H_{123}-3H_{125}}{4}M_LM_T^2 \\
&\qquad\qquad\left.+\frac{\Delta H\sin(4\alpha)}{4}\left(M_T^3-3M_L^2M_T\right) \right].
\end{split}
\label{eq:Kerr001_final}
\end{equation}
  
\noindent
As in this study only in-plane magnetization is investigated, we set $M_P=0$ to derive the final equation for (001)-oriented cubic crystal structures. Nevertheless, note that if an $M_P$ contribution was present, only the offset of the MOKE signal would be affected since Eqs.~(\ref{eq:eps_xy001}-\ref{eq:eps_yz001}) do not include any $M_P$ contributions that are connected to an angular dependence on $\alpha$. Therefore, no additional anisotropy upon sample rotation would be induced. 

\subsection{Analytical equations of MOKE for (111)-oriented cubic crystal structures}
\label{sec:Kerr111_final}

To provide the analogous analytical equations for (111)-oriented cubic crystal structures, the anisotropic MO tensors $\bm{G}$ and $\bm{H}$,  must be transformed into their (111)-oriented forms. We define the transformation
\begin{equation}
	\bm{a}^{(001)\rightarrow (111)}=
	\begin{bmatrix}
		-\frac{\sqrt{6}}{3}&\frac{\sqrt{6}}{6}&\frac{\sqrt{6}}{6}\\
		0& -\frac{\sqrt{2}}{2}& \frac{\sqrt{2}}{2} \\
		\frac{\sqrt{3}}{3}&\frac{\sqrt{3}}{3} & \frac{\sqrt{3}}{3} \\
	\end{bmatrix}
	\label{eq:a_rot_to_111}
\end{equation}

\noindent
which, through  Eqs.~\eqref{eq:g_rot} and \eqref{eq:h_rot}, yields the $\bm{G}$ and $\bm{H}$ tensors for (111)-oriented cubic crystal structures in the reference orientation.  Finally, we apply a rotation about the $\hat{z}$-axis by a general angle $\alpha$ to describe the rotation of the sample around its surface normal. See appendix \ref{app:conventions} for more details, and its Eq.\ref{eq:a_rot_z} for definition of the rotation matrix about $\hat{z}$-axis. 

The off-diagonal elements of the permittivity tensor of (111)-oriented cubic crystal structures described up to third order in $\bm{M}$ and for a general sample orientation $\alpha$ are 

\begin{equation}
\begin{split}
\varepsilon^{(111)}_{xy/yx}= \qquad \qquad \qquad \pm K\quad & M_P\\
+\frac{\sqrt{2}}{3}\Delta G \cos(3\alpha)\quad & M_LM_P\\
-\frac{\sqrt{2}}{3}\Delta G \sin(3\alpha)\quad & M_PM_T\\
+\left(2G_{44}+\frac{1}{3}\Delta G\right)\quad &M_LM_T \\
\mp\frac{1}{3\sqrt{2}}\Delta H\cos(3\alpha)\quad &M_T^{3}\\
\pm\frac{1}{3\sqrt{2}}\Delta H\sin(3\alpha)\quad &M_L^{3}\\
\pm \frac{1}{3}\left(H_{123}+6H_{125}\right) \quad & M_P^3 \\
\pm H_{123}\quad & M_PM_T^2\\
\pm\frac{1}{\sqrt{2}}\Delta H\cos(3\alpha)\quad &M_TM_L^{2}\\
\pm H_{123}\quad & M_PM_L^2\\
\mp\frac{1}{\sqrt{2}}\Delta H\sin(3\alpha)\quad &M_LM_T^{2}\\
\end{split} 
\end{equation}

\begin{equation}
\begin{split}
\varepsilon^{(111)}_{zx/xz}=\qquad \qquad \qquad \pm K\quad &M_L\\
+\frac{1}{3}(2G_s+2G_{44}) \quad & M_PM_T\\
-\frac{2}{3\sqrt{2}}\Delta G \sin{(3\alpha)}\quad &M_LM_T\\
-\frac{1}{3\sqrt{2}}\Delta G \cos{(3\alpha)}\quad &(M_T^2-M_L^2)\\
\pm\frac{1}{2}\left(H_{123}+3H_{125}\right)\quad &M_L^3\\
\pm H_{123} \quad & M_LM_P^2\\
\mp \frac{\sqrt{2}}{2}\Delta H\sin(3\alpha) \quad &M_PM_T^2\\
\pm \frac{\sqrt{2}}{2}\Delta H\sin(3\alpha) \quad &M_PM_L^2\\
\pm\frac{1}{2}\left(H_{123}+3H_{125}\right)\quad &M_LM_T^2\\
\pm \sqrt{2}\Delta H\cos(3\alpha)\quad & M_TM_LM_P
\end{split}
\end{equation}
and

\begin{equation}
\begin{split}
\varepsilon^{(111)}_{yz/zy}=\qquad \qquad \qquad\pm K \quad &M_T\\
+\frac{1}{3}(2G_s+2G_{     44})\quad &M_LM_P\\
+ \frac{2}{3\sqrt{2}}\Delta G \cos{(3\alpha)}\quad &M_LM_T\\
- \frac{1}{3\sqrt{2}}\Delta G \sin{(3\alpha)}\quad &(M_T^2-M_L^2)\\
\pm \frac{1}{2}\left(H_{123}+3H_{125}\right)\quad &M_T^3\\
\mp \frac{\sqrt{2}}{2}\Delta H\cos(3\alpha)\quad & M_PM_T^2\\
\pm \frac{1}{2}\left(H_{123}+3H_{125}\right)\quad & M_TM_L^2\\
\pm \frac{\sqrt{2}}{2}\Delta H\cos(3\alpha)\quad & M_PM_L^2\\
\pm H_{123}\quad & M_TM_P^2\\
\mp \sqrt{2}\Delta H\sin(3\alpha)\quad & M_TM_LM_P.
\end{split}
\end{equation}
As the cross-term of type $\varepsilon_{yz/zy}\, \varepsilon_{zx/xz}/\varepsilon_d$ that is present in Eq.~\eqref{Kerr_analyt} is of more complicated nature for (111)-oriented cubic crystal structures then in case of (001)-oriented cubic crystal structures, we evaluate its shape here as well.
\begin{equation}
\begin{split}
\varepsilon_{yz/zy}^{(111)} \varepsilon_{zx/xz}^{(111)}=\qquad \qquad \qquad  \quad K^2\quad &M_LM_T\\
 \mp\frac{1}{3\sqrt{2}}K\Delta G \cos{(3\alpha)}\quad &M_T^3\\
 \pm\frac{1}{3\sqrt{2}}K\Delta G \sin{(3\alpha)}\quad &M_L^3\\
 \pm \frac{1}{3}(2KG_s+K2G_{44})\quad & M_PM_T^2\\
 \pm\frac{1}{\sqrt{2}}K\Delta G \cos{(3\alpha)}\quad &M_TM_L^2\\
   \pm \frac{1}{3}(2KG_s+K2G_{44})\quad & M_PM_L^2\\
 \mp\frac{1}{\sqrt{2}}K\Delta G \sin{(3\alpha)}\quad &M_LM_T^2.
\end{split}
\end{equation}

Note that the contributions of 4th and higher orders are neglected in this cross-term. Using these elements in Eq.~\eqref{Kerr_analyt} will provide us with an analytical expression of the Kerr effect dependence on the MO parameters, in-plane magnetization components and sample orientation $\alpha$ for (111)-oriented thin films with cubic crystal structure. We derive

\begin{widetext}
\begin{equation}
\begin{split}	
\Phi_{s/p, [M_P=0]}^{(111)} =& \\
& A_{s/p} \left[\pm \left(2G_{44}+\frac{\Delta G}{3}-\frac{K^2}{\varepsilon_d}\right)M_LM_T 
-\frac{1}{3\sqrt{2}}\left(\Delta H+\frac{K\Delta G}{\varepsilon_d}\right)\sin(3\alpha)\left(M_L^{3}-3M_LM_T^2\right)\right. \\
& \qquad \qquad \qquad \qquad \qquad \left. +\frac{1}{3\sqrt{2}}\left(\Delta H+\frac{K\Delta G}{\varepsilon_d}\right)\cos(3\alpha)\left(M_T^{3}-3M_TM_L^2\right)\right]
\\[5mm]
+&B_{s/p} \left[\pm KM_L-\frac{2\Delta G}{3\sqrt{2}}\sin{(3\alpha)}M_LM_T-\frac{\Delta G}{3\sqrt{2}}\cos{(3\alpha)}(M_T^2-M_L^2)\pm\frac{H_{123}+3H_{125}}{2}\left(M_L^3+M_LM_T^2\right)\right].
\end{split}
\label{eq:Kerr111_final}
\end{equation}
\end{widetext}

Again we restricted the equation to in-plane magnetization only and set $M_P=0$. Yet, Eq.~\eqref{eq:Kerr111_final} can be easily extended to all magnetization components using all the equations presented above.  If $M_P$ component were included, it would generate, in addition to offset, also new anisotropies starting already from second order in magnetization, being namely $\pm A_{s/p}[-(\sqrt{2}/3) \Delta G \cos{(3\alpha)}M_LM_P +(\sqrt{2}/3) \Delta G \sin{(3\alpha)}M_TM_P]$. Note that this is in contrast to the case of (001) orientation in  Eq.~\eqref{eq:Kerr001_final}, for which no such angular dependencies would be created.

\subsection{Eight-directional method measurements}

The so-called eight-directional method \cite{Postava2002} is a measurement algorithm, that was designed to separate LinMOKE and QMOKE contributions and to identify their angular dependencies, but, as shown here, it can also be used to investigate CMOKE effects. At each sample orientation $\alpha$ the MOKE signal $\Phi^{(\mu)}$ is detected for eight in-plane magnetization directions  $\mu=\{ 0^\circ,45^\circ,90^\circ, \ldots, 270^\circ,315^\circ\}$. Proper linear combinations of those signals remove the experimental MOKE background. We use four different linear combinations of those signals to separate four experimental MOKE contributions. The first contribution that is separated is
\begin{equation}
 \Phi_{M_L, M_L^3}=  \frac{\Phi_{s/p}^{\mu = 90^\circ } - \Phi_{s/p}^{\mu = 270^\circ}}{2}
 \label{eq:LCMOKE_measure}
\end{equation}
being the MOKE signal that combines the LMOKE$\sim M_L$ and longitudinal CMOKE (LCMOKE)$\sim M_L^3$ contributions. The second contribution is 
\begin{equation}
\Phi_{M_LM_T}=\frac{\Phi^{\mu = 45^\circ } + \Phi^{\mu = 225^\circ } - \Phi^{\mu = 135^\circ } - \Phi^{\mu = 315^\circ }}{2}
	\label{eq:QMOKE_MLMT_measure}
\end{equation}
being the QMOKE$\sim M_LM_T$ contribution. The third contribution is 
\begin{equation}
	\Phi_{M_T^2-M_L^2}= \frac{\Phi^{\mu = 0^\circ } + \Phi^{\mu = 180^\circ } - \Phi^{\mu = 90^\circ } - \Phi^{\mu = 270^\circ }}{2}
	\label{eq:QMOKE_MT2MT2_measure}
\end{equation}
being the QMOKE$\sim (M_T^2-M_L^2)$ contribution. Finally, the fourth contribution is 
\begin{equation}
 \Phi_{M_T^3}=  \frac{\Phi^{\mu = 0^\circ } - \Phi^{\mu = 180^\circ}}{2} 
 \label{eq:TCMOKE_measure}
\end{equation}
being the transversal CMOKE (TCMOKE)$\sim M_T^3$ contribution.

\begin{center}
\begin{table*}
\begin{tabular}{lrcrc}
			\hline
			\hline	
			Contribution && offset && angular dependence \\
			\hline \\
			 \makecell{LMOKE+LCMOKE \\ $\Phi_{M_L,M_L^3}$} && $\pm B_{s/p} \left[ K + \frac{3}{4}(H_{123}+H_{125})\right] $   && $\pm B_{s/p}\frac{1}{4}\Delta H\cos(4\alpha)$  \\[5mm]
			 \makecell{TCMOKE \\$\Phi_{M_T^3}$	 }	 && -- && $\pm B_{s/p}\frac{1}{4}\Delta H\sin(4\alpha)$  \\[5mm]
			  \makecell{QMOKE \\$\Phi_{M_LM_T}$}	 &&  $\pm A_{s/p} \left(2G_{44}+\frac{1}{2}\Delta G-K^2/\varepsilon_d\right)$  && $\mp A_{s/p}\frac{1}{2}\Delta G\cos{(4\alpha)}$ \\[5mm]
			 \makecell{QMOKE \\$\Phi_{M_T^2-M_L^2}$}	 &&  --   && $\pm A_{s/p}\frac{1}{2}\Delta G\sin{(4\alpha)}$ \\[3
			 mm]
		\hline
		\hline
\end{tabular}
\caption{Dependencies of contributions separated by eight-directional method (i.e. according to Eq.~\eqref{eq:Kerr001_final} and Eqs. (\protect{\ref{eq:LCMOKE_measure}-\ref{eq:TCMOKE_measure}})) on MO parameters and sample orientation angle $\alpha$ for (001)-oriented cubic crystal structures.}
\label{tab:sample_Ni001_contributions}
\end{table*}
\end{center}

\begin{center}
\begin{table*}
\begin{tabular}{lrcrc}
			\hline
			\hline	
			Contribution && offset && angular dependence \\
			\hline \\
			  \makecell{LMOKE+LCMOKE \\ $\Phi_{M_L,M_L^3}$} && $\pm B_{s/p} \left[ K + \frac{1}{2}(H_{123}+3H_{125})\right] $   && $-A_{s/p}\,\frac{\sqrt{2}}{6}\left(\Delta H+K\Delta G/\varepsilon_d\right)\sin(3\alpha)$  \\[5mm]
			 \makecell{TCMOKE \\$\Phi_{M_T^3}$	 }&& -- && $A_{s/p}\,\frac{\sqrt{2}}{6}\left(\Delta H+K\Delta G/\varepsilon_d\right)\cos(3\alpha)$  \\[5mm]
			 \makecell{QMOKE \\$\Phi_{M_LM_T}$}	 &&  $\pm A_{s/p}\left(2G_{44}+\frac{1}{3}\Delta G-K^2/\varepsilon_d\right)$  && $-B_{s/p}\,\frac{\sqrt{2}}{3}\Delta G\sin{(3\alpha)}$ \\[5mm]
			 \makecell{QMOKE \\$\Phi_{M_T^2-M_L^2}$}	 &&  --   && $-B_{s/p}\,\frac{\sqrt{2}}{3}\Delta G\cos{(3\alpha)}$ \\[3
			 mm]
		\hline
		\hline
\end{tabular}
\caption{Dependencies of contributions separated by the eight-directional method (i.e. according to Eq.~\eqref{eq:Kerr111_final} and Eqs. (\protect{\ref{eq:LCMOKE_measure} -\ref{eq:TCMOKE_measure}})) on MO parameters and sample orientation angle $\alpha$ for (111)-oriented cubic crystal structures. Note that the sign of the angular-dependence contributions depends on the choice of the reference in-plane orientation, $\alpha=0^{\circ}$, for the (111)-oriented cubic crystal structure. A phase shift of $60^\circ$ or $180^\circ$ relative to our reference orientation (described in Appendix~\protect{\ref{app:conventions}}) will reverse the sign.}
\label{tab:sample_Ni111_contributions}
\end{table*}
\end{center}

Substituting the expressions for $\Phi^{(\mu)}$ from  Eq.~\eqref{eq:Kerr001_final} and Eqs. (\ref{eq:LCMOKE_measure}-\ref{eq:TCMOKE_measure}), we can describe analytical expressions of individual contributions separated by the eight-directional method for (001)-oriented cubic crystal structures. In Tab.~\ref{tab:sample_Ni001_contributions}, we show each contribution separated by its offset and angular dependence. Here, we can see that with normal AoI only QMOKE contributions, which are proportional to the optical weighting factor $A_{s/p}$ will persist. Namely, these are the $\Phi_{M_LM_T}$  contribution which will possess a finite offset as well as a four-fold angular dependence of $\cos(4\alpha)$ and $\Phi_{M_T^2-M_L^2}$ which only consists of an angular dependence of $\sin(4\alpha)$. The amplitudes of the angular dependencies of $\Phi_{M_LM_T}$ and $\Phi_{M_T^2-M_L^2}$ are identical. With oblique AoI, also contributions weighted by optical factor $B_{s/p}$ will emerge, being namely the LMOKE+LCMOKE contribution $\Phi_{M_L,M_L^3}$ with finite offset and an angular dependence of $\cos(4\alpha)$, and the TCMOKE contribution $\Phi_{M_T^3}$ with an angular dependence of $\sin(4\alpha)$ only. The amplitudes of the angular dependencies of $\Phi_{M_L,M_L^3}$ and $\Phi_{M_T^3}$ are again the same. Finally, note that for (001)-oriented cubic crystal structures, all contributions separated by the eight-directional method change their sign when the polarization of the incident light is changed.

In the case of (111)-oriented cubic crystal structures, we can describe individual contributions to the eight-directional method measurement by using Eq.~\eqref{eq:Kerr111_final} with Eqs. (\ref{eq:LCMOKE_measure}-\ref{eq:TCMOKE_measure}). The main differences with respect to (001)-oriented cubic crystal structures are now the three-fold angular dependencies of contributions separated by the eight-directional method and their different dependence on the optical weighting factors $A_{s/p}$ and $B_{s/p}$, as can be seen in Tab \ref{tab:sample_Ni111_contributions}. The three-fold angular dependence of $\sin(3\alpha)$ with the QMOKE contributions $\Phi_{M_LM_T}$ and of $\cos(3\alpha)$ with the QMOKE contribution $\Phi_{M_T^2-M_L^2}$  will only appear with oblique AoI due to its proportionality to $B_{s/p}$. The amplitude of these two angular dependencies is again the same. For the LMOKE+LCMOKE and TCMOKE contributions, we observe the angular dependencies $\sin(3\alpha)$ for LMOKE+LCMOKE $\Phi_{M_L,M_L^3}$ and $\cos(3\alpha)$ for TCMOKE $\Phi_{M_T^3}$ already at normal AoI, since they are proportional to $A_{s/p}$. Again the amplitude of these two angular dependencies is the same. Interestingly, the signs of the amplitudes of the angular dependencies are invariant to the change of the polarization of the incident light now. The offset of LMOKE+LCMOKE $\Phi_{M_L,M_L^3}$ and offset of QMOKE $\Phi_{M_LM_T}$ have the same dependence on optical weighting factors as in the case of (001)-oriented cubic structures. The change of the sign with the change of the polarization of incident light is also preserved. It is only the value of the offset that is changed from the case of (001)-oriented cubic crystal structures, where MO parameters $H_{123}$, $H_{125}$ and $\Delta G$ contribute by different amounts.

Note that the eight-directional method cannot separate the LCMOKE contribution $\sim M_L^3$ from the LMOKE $\sim M_L$ contribution and thus these two contributions are superimposed in our experimental data. However, the LCMOKE contribution solely determines the amplitude of the angular dependence of the $\Phi_{M_L,M_L^3}$ contribution for (001)-oriented cubic crystal structures and dominates the amplitude of the angular dependence of the $\Phi_{M_L,M_L^3}$ contribution for (111)-oriented cubic crystal structures. The latter has been proven by numerical simulations done in the last part of our prior CMOKE publication \cite{Gaerner24}. Thus, LCMOKE can be separated by a combination of the eight-directional method and the analysis of the obtained angular dependence of the $\Phi_{M_L,M_L^3}$ contribution. Furthermore, in Eqs.~\eqref{eq:Kerr001_final} and \eqref{eq:Kerr111_final} there are contributions proportional to $M_T^2M_L$ and $M_TM_L^2$. However, these contributions are being filtered out in all MOKE signals separated by the eight-directional method according to Eqs. (\ref{eq:LCMOKE_measure}-\ref{eq:TCMOKE_measure}). Finally, from Eq.~\eqref{eq:Kerr001_final} and Eq.~\eqref{eq:Kerr111_final} it seems that the QMOKE angular dependence of $\Phi_{M_LM_T}$ should have twice the amplitude compared to the QMOKE angular dependence $\Phi_{M_T^2-M_L^2}$. But note that the product $M_LM_T$ of the normalized magnetization components  $M_L$ and $M_T$, when being measured with magnetization alignment along the 45$^\circ$ directions as described in Eq.~\eqref{eq:QMOKE_MLMT_measure}, will effectively produce a factor $\sin(\pi/4)\cdot \cos(\pi/4)=1/\sqrt(2)\cdot 1/\sqrt(2)=1/2$.  Thus, the amplitudes of the angular dependencies of $\Phi_{M_LM_T}$ and $\Phi_{M_T^2-M_L^2}$ are eventually the same here.

%%%%%%%%%%%%%%%%%%%%%%%%%%%%%%%%%%%%%%%%%%%
%---Experimental observations
%%%%%%%%%%%%%%%%%%%%%%%%%%%%%%%%%%%%%%%%%%%
\section{Experimental observations and numerical model fitting}
\label{experimental observations}
\subsection{Sample description and experimental background}

In this section, we present experimental measurements of two samples using the eight-directional method, being sample~Ni(111): MgO(111)// Ni(22.5\,nm)/ Si(2.2\,nm)/ SiO$_x$(0.9\,nm), and sample~Ni(001): MgO(001)// Ni(18.8\,nm)/ Pt(2.2\,nm). For the structural characterization of sample~Ni(111), see Ref. \cite{Gaerner24}. For the structural characterization of sample~Ni(001), see the Appendix \ref{app:struct_char_001}. In short, from x-ray diffraction (XRD), including specular $\Theta-2\Theta$ scans and off-specular Euler cradle measurements, it is shown that the Ni layers in both samples are well epitaxially grown and that the Ni(111) and Ni(001) layers are (111) and (001) oriented, respectively. Since Ni prefers face-centered cubic crystal growth, two in-plane growth orientations of (111)-oriented cubic crystal Ni on MgO(111) are possible. In fact, a small twinning effect is present in the Ni(111) sample, but the dominant phase makes up roughly 95$\%$ of the total intensity in off-specular Euler cradle x-ray scans \cite{Gaerner24}. 

The eight-directional method measurements of both samples were carried out using a setup allowing measurements of the MOKE response under 0$^\circ$ or 45$^\circ$ AoI and for all eight in-plane directions of the magnetic field as specified in Eqs.~(\ref{eq:LCMOKE_measure}-\ref{eq:TCMOKE_measure}). The setup can reach  magnetic field strengths up to 235\,mT, which is enough to achieve saturation for all eight in-plane directions with both samples. To acquire angular dependencies of the experimentally separated contributions, the setup allows for an in-plane rotation of the sample. The proper alignment of the sample is possible through a rotatable and tiltable sample holder stage. The eight-directional method measurements were performed using 3$^\circ$ sample rotation steps within all measurements in this study. As a light source we use a fibre-coupled laser source where the laser diodes wavelengths are: 635\,nm $\approx$ 1.95\,eV and 406\,nm $\approx$ 3.05\,eV. For additional information on the MOKE experimental setup, see Refs. \cite{Kehlberger15, Muglich16}.

The experimental results of the eight-directional method are compared to the analytical expressions in Tab. \ref{tab:sample_Ni001_contributions} and Tab. \ref{tab:sample_Ni111_contributions}, from which the amplitude and offset of each separated contribution is evaluated. To be able to effectively describe and simulate the experimental data starting from MO parameters, an in-house developed python code based on Yeh's  4$\times$4 transfer matrix formalism \cite{Yeh1980} has been used.  The  model is based on coherent propagation of a monochromatic plane wave in the multilayer structures, using polarization eigenmodes and Maxwell's boundary conditions. The input parameters for the model are \textit{(i)} thickness (gathered by x-ray reflectivity (XRR) - see Appendix~\ref{app:struct_char_001} and Ref. \cite{Gaerner24}) and \textit{(ii)} permittivity (gathered by ellipsometry - see Appendix \ref{app:ellipsometry} and Ref. \cite{Gaerner24}) of each layer in the stack. Further, \textit{(iii)} AoI and \textit{(iv)} wavelength of the incident light are directly known from the experiment. The reflection coefficients (and thus the Kerr angles) of the sample can be calculated for arbitrary \textit{(v)} magnetization direction and \textit{(vi)} orientation $\alpha$ of the sample (both also being known from the experiment), whereas \textit{(vii)} the MO parameters $K$, $G_s$, $2G_{44}$, and $\Delta H$ are set as free parameters of the model. This allows us to numerically reproduce the eight-directional method measurements within the boundaries of classical electromagnetic wave optics. The MO parameters are then yielded by fitting this model to the experimental data.

Nevertheless, in the fitting process, we are unable to find the value of all MO parameters independently. First, the offset of LMOKE+LCMOKE, for both (001)- and (111)-oriented cubic crystal structures, is given by the MO parameters $K, H_{123}$ and $3H_{125}$ and we are not able to evaluate their contribution to this offset independently. Secondly, although value of $\Delta G$ is provided by the amplitude of the QMOKE $\Phi_{M_LM_T}$ and QMOKE $\Phi_{M_T^2-M_L^2}$ angular dependencies, the contribution of $K\Delta G/\varepsilon_d$ to the amplitude of the LCMOKE and TCMOKE angular dependencies (in case of (111)-oriented cubic crystal structure) cannot be exactly evaluated, as we do not posses exact information on the value of $K$. For (001)-oriented cubic crystal structures, the value of $\Delta H$ can be evaluated from the angular dependencies of LCMOKE and TCMOKE, but as $\Delta H = H_{123}-3H_{125}$, we are not able to determine the specific values of $H_{123}$ and $3H_{125}$ alone. For clarity see Tab. \ref{tab:sample_Ni001_contributions} and Tab. \ref{tab:sample_Ni111_contributions}. Although above reasoning is based on analytical expressions which stem from Eq.~\eqref{Kerr_analyt} and which the numerical model does not use, we observe this issue through the complete correlation of $K, H_{123}$ and $3H_{125}$ in the numerical model. Therefore, to resolve this issue, in our fitting routine we adopted the policy to minimize the isotropic contribution of CMOKE, (i.e. of MO tensor $\bm{H}$) when describing the experimental data. This is done by keeping $3H_{125}=0$, thus $\Delta H=H_{123}$. 

Finally, note that although it may seem that, from a qualitative point of view, the contribution of the MO tensor $\bm{H}$ is not needed at all for (111)-oriented cubic crystal structures (see that the offset of LMOKE+LCMOKE could be described solely by $K$, whereas the angular dependencies of LCMOKE and TCMOKE could be described solely by $K\Delta G/\varepsilon_d$), quantitatively the contribution of $\bm{H}$ is indispensable to describe the experimental data of sample Ni(111). Without $\bm{H}$ the value of $K$ is given by the offset of LMOKE+LCMOKE, and the value of $\Delta G$ is given by the QMOKE angular dependencies. Then, the contribution of $K\Delta G/\varepsilon_d$ to the LCMOKE and TCMOKE angular dependencies is rather negligible and cannot describe the experimental data alone, as we have proven in our prior CMOKE paper \cite{Gaerner24}.

%%%%%%%%%%%%%%%%%%%%%%%%%%%%%%%%%%%%%%%%%%%
%---Experimental Ni (111)
%%%%%%%%%%%%%%%%%%%%%%%%%%%%%%%%%%%%%%%%%%%
\subsection{Eight-directional method measurement of the Ni(111) sample}

\begin{figure*}
\includegraphics{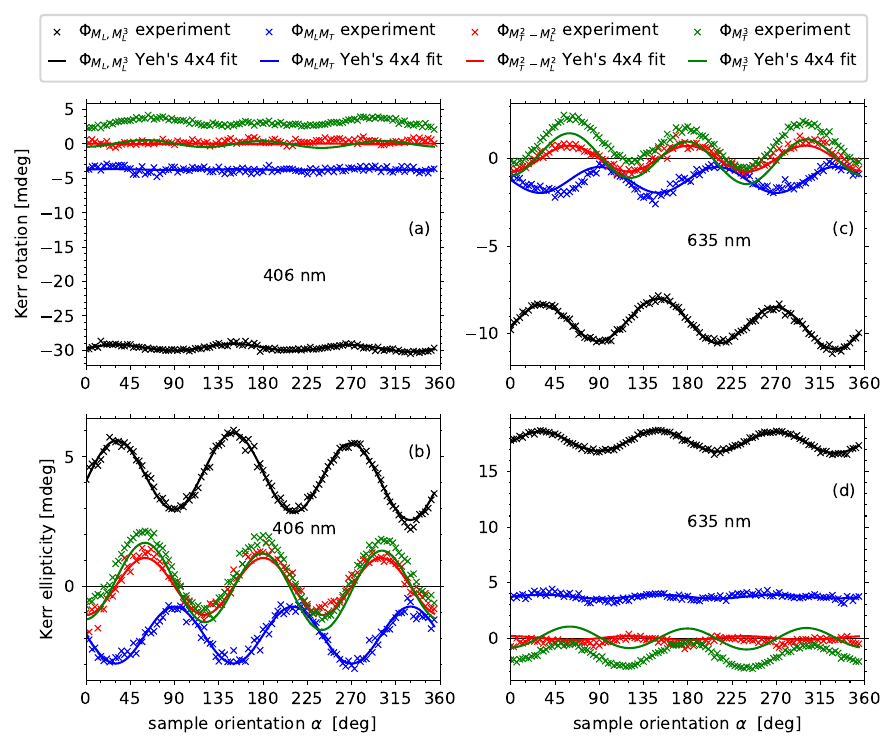}
\caption{Eight-directional method measurement of sample~Ni(111) with $p$-polarized light and at an AoI of 45$^\circ$. Graphs present the dependence on the in-plane sample orientation $\alpha$ of the (a) Kerr rotation and (b) Kerr ellipticity at a wavelength of 406\,nm as well as (c) Kerr rotation and (d) Kerr ellipticity at a wavelength of 635\,nm. Experimental data points are plotted as "x" marks, whereas solid lines represent the fit of the numerical model based on Yeh's 4$\times$4 transfer matrix formalism to the experimental data. }
\label{f:8dir_111RS}
\end{figure*}

%TABLE AMPLITUDES AND OFFSETS FROM RS 111
\begin{center}
\begin{table}
\begin{tabular}{lrcrcrcrc}
			\hline
			\hline
			&&\multicolumn{7}{c}{Kerr rotation [mdeg]}\\
				\hline
				 && \multicolumn{3}{c}{406\,nm} && \multicolumn{3}{c}{635\,nm}   \\
				 \hline
			       && offset &&  \makecell{3-fold \\  amplitude}  && offset &&  \makecell{3-fold \\  amplitude} \\

			\hline
			 $M_L,M_L^3$	 && -29.65$\pm$0.02  && 0.47$\pm$0.02 && -9.45$\pm$0.01 && 1.17$\pm$0.01 \\[1mm]
			 $M_T^3$	 && 3.13$\pm$0.03  && -0.43$\pm$0.04 &&  0.86$\pm$0.02 && -1.16$\pm$0.02 \\[1mm]
			 $M_LM_T$	 &&  -3.73$\pm$0.03  && 0.10$\pm$0.05 &&  -1.22$\pm$0.02 &&  -0.71$\pm$0.03\\[1mm]
			 $M_T^2-M_L^2$	 &&  0.32$\pm$0.03   && 0.02$\pm$0.04 && 0.14$\pm$0.02 && -0.77$\pm$0.02 \\[1mm]
			 
			 			\hline
			&&\multicolumn{7}{c}{Kerr ellipticity [mdeg]}\\
				\hline
				 && \multicolumn{3}{c}{406\,nm} && \multicolumn{3}{c}{635\,nm}   \\
				 \hline
			       && offset &&  \makecell{3-fold \\  amplitude}  && offset &&  \makecell{3-fold \\  amplitude} \\

			\hline
			 $M_L,M_L^3$	 && 4.24$\pm$0.01  && 1.44$\pm$0.02 && 17.71$\pm$0.01 && 0.95$\pm$0.01 \\[1mm]
			 $M_T^3$	 && 0.48$\pm$0.01  && -1.44$\pm$0.02 &&  -1.49$\pm$0.02 && -0.94$\pm$0.03 \\[1mm]
			 $M_LM_T$	 &&  -1.87$\pm$0.02  && -1.03$\pm$0.02 &&  3.73$\pm$0.02 &&  0.23$\pm$0.03\\[1mm]
			 $M_T^2-M_L^2$	 &&  0.16$\pm$0.02   && -1.16$\pm$0.03 && -0.26$\pm$0.02 && 0.12$\pm$0.03 \\[1mm]
		\hline
		\hline
\end{tabular}
\caption{Offset values and 3-fold amplitudes of contributions experimentally measured by the eight-directional method on sample~Ni(111) as presented in  Fig. \protect\ref{f:8dir_111RS}.}
\label{tab:RS111_amplitude_offset}
\end{table}
\end{center}

% TABLE MO PARAMETERS FROM RS 111
\begin{center}
\begin{table*}
\begin{tabular}{lrcrc}
			\hline
			\hline
			MO parameter      && 406 nm && 635 nm  \\

			\hline\\
			 $K$	 &&  -0.0402$\pm$0.0005\,-(0.1161$\pm$0.0005)i  &&-0.2490$\pm$0.0008\,+(0.0114$\pm$0.0008)i  \\[1mm]
			 $G_s$	 && 0.0041$\pm$0.0006\,-(0.0017$\pm$0.0006)i  && -0.0092$\pm$0.0010\,-(0.0054$\pm$0.0010)i  \\[1mm]
			 $2G_{44}$	 &&  -0.0052$\pm$0.0003\,-(0.0000(4)$\pm$0.0003)i  && 0.0044$\pm$0.0005\,+(0.0087$\pm$0.0005)i\\[1mm]
			 $\Delta G$	 &&  0.0093$\pm$0.0009\,-(0.0017$\pm$0.0009)i  && -0.0136$\pm$0.0015\,-(0.0141$\pm$0.0015)i\\[1mm]
			 $\Delta H$	 &&  0.0023$\pm$0.0003\,+(0.0046$\pm$0.0003)i   && 0.0100$\pm$0.0004\,+(0.0018$\pm$0.0004)i \\[1mm]
			 \hline
			 vicinal parameters      &&  &&   \\
			 \hline
			 $\varepsilon_S$	 &&  0.0114$\pm$0.0096\,-(0.0203$\pm$0.0096)i  &&  0.0162$\pm$0.0093\,-(0.0439$\pm$0.0093)i \\[1mm]
		$\alpha_1$ && -46$^\circ$$\pm$23$^\circ$ && -46$^\circ$$\pm$11 $^\circ$ \\
%		$\Delta\alpha$ && 0$^\circ$ && 0 $^\circ$ \\
		\hline
		\hline
\end{tabular}
\caption{Values of MO parameters for Ni layer of sample~Ni(111), provided by the fit of Yeh's 4$\times$4 transfer matrix numerical model to the experimental data as is shown in Fig. \protect\ref{f:8dir_111RS}. The value of $\Delta G = G_s-2G_{44}$ is shown for comparison with $\Delta H$. The values of $\varepsilon_S$ and $\alpha_1$ are related to VISMOKE as described in Appendix \ref{app:vismoke}.  }
\label{tab:RS111_fit_result}
\end{table*}
\end{center}

In Fig. \ref{f:8dir_111RS}, we show experimental results of the eight-directional method measurement for the sample~Ni(111). Individual contributions to the Kerr rotation and Kerr ellipticity separated by the eight-directional method are plotted for a wavelength of 406\,nm in Figs. \ref{f:8dir_111RS}(a) and \ref{f:8dir_111RS}(b), respectively, whereas they are plotted for a wavelength of 635\,nm in Figs. \ref{f:8dir_111RS}(c) and \ref{f:8dir_111RS}(d), respectively. All these measurements were executed with $p$-polarized incident light and at an AoI of 45$^\circ$. The contributions in each graph of Fig. \ref{f:8dir_111RS}(a)-(d) are measured according to Eqs. (\ref{eq:LCMOKE_measure}-\ref{eq:TCMOKE_measure}). To compare with analytical expressions of those contributions see Tab.~\ref{tab:sample_Ni111_contributions}. 

It is clearly visible that most of the experimental findings in Fig. \ref{f:8dir_111RS} follow the predictions of the analytical model very well, with only two exceptions. (i) The one-fold contribution, which is present in all contributions separated by the eight-directional method, but being mostly pronounced with LMOKE+LCMOKE and TCMOKE contributions, is not described by Eq.~\eqref{eq:Kerr111_final} and thus not shown in Tab.~\ref{tab:sample_Ni111_contributions}. This one-fold contribution is stemming from the interplay of a vicinal surface and magneto-optics, thus it is named  vicinal interface sensitive MOKE (VISMOKE)\cite{Hamrle03}. Note, that although not shown in the analytical formulas, the numerical model does account for this VISMOKE contribution as is discussed below and in Appendix \ref{app:vismoke}. (ii) The offset of the TCMOKE $\sim M_T^3$ contribution present in the experimental data is not predicted by the analytical model, and the numerical model is unable to describe it as well. So far, we excluded some slight external magnetic field misalignment in the experimental setup as a possible source of this offset \cite{Gaerner24} and further investigations are needed to conclude on the origin of this contribution.
 
To evaluate amplitude and offset of each contribution, we first describe the data using a simple equation
\begin{equation}
	\Phi_x(\alpha)=A_3\sin(3(\alpha-\alpha_3))+A_1\sin(\alpha-\alpha_1)+C.
	\label{eq:simple_model_111}
\end{equation}

\noindent
Here $x=\{(M_L,M_L^3), M_T^3, M_LM_T,(M_T^2-M_L^2)\}$, the offset is described by $C$, the amplitude of the three-fold angular dependence is described by $A_3$ and the phase shift $\alpha_3$ describes a possible misalignment of the sample in the setup (found to be negligible) and/or allows to describe the contributions with the $\cos(3\alpha)$ dependence by providing $90^\circ$ phase shift. The amplitude $A_1$ of the one-fold angular dependence and the phase shift $\alpha_1$ are stemming from VISMOKE and are not within the focus of this paper. The correlation between the phase shift $\alpha_1$ and the vicinal direction of the surface detected by XRD has already been discussed in our prior work \cite{Gaerner24}. In Tab. \ref{tab:RS111_amplitude_offset}, offset values and amplitudes are presented. Note that the absolute amplitude of LCMOKE compared to the absolute amplitude of TCMOKE is almost the same for all four cases (406nm / 635nm, Kerr rotation / Kerr ellipticity) presented in Fig. \ref{f:8dir_111RS} as predicted by the theory. The signs also follow the prediction shown in Tab. \ref{tab:sample_Ni111_contributions}. The amplitude and sign of the contributions QMOKE$\sim M_LM_T$ and QMOKE$\sim (M_T^2-M_L^2)$ also follow the predictions of Tab. \ref{tab:sample_Ni111_contributions} very well, except for the case of Fig. \ref{f:8dir_111RS}(a) and (d), where the amplitude of these angular dependencies is near zero, thus the signal-to-noise ratio is low, resulting in a poor fit and slight differences in the absolute values of the amplitude here.

The solid lines in Fig.~\ref{f:8dir_111RS} represent the fit of the Yeh's 4$\times$4 transfer matrix model to the experimental data. The MO parameters $K,G_s,2G_{44}$ and $ \Delta H$ together with the parameters $\varepsilon_S$ and $\alpha_1$ of the vicinal surface (see Appendix \ref{app:vismoke}) were set as free parameters. The values provided by the fit are displayed in Tab. \ref{tab:RS111_fit_result}. Here we can see that, in absolute values, the anisotropy strength of the MO tensor $\bm{H}$ expressed by $\Delta H$ is roughly one half of the anisotropy strength of the MO tensor $\bm{G}$ expressed by $\Delta G$. Yet from Tab. \ref{tab:RS111_amplitude_offset} we can see that the amplitude of the CMOKE angular dependence is rather a bit stronger than the amplitude of the QMOKE angular dependence. The reason is that $\Delta G$ is weighted by $B_{p}$ whereas $\Delta H$ is weighted by $A_{p}$, while $A_{s/p}$ tends to be stronger than $B_{s/p}$. Consider that for the LinMOKE approximation LMOKE=$\pm B_{s/p}KM_L$ and PMOKE=$A_{s/p}KM_P$, while PMOKE is generally stronger than LMOKE \cite{Visnovsky06}, due to the generally stronger $A_{s/p}$ parameter compared to $B_{s/p}$.

\begin{figure}
\includegraphics{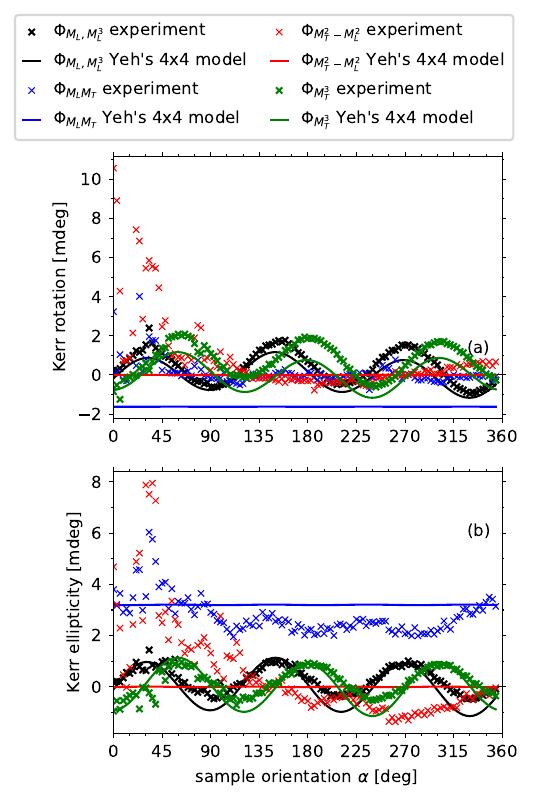}
\caption{Eight-directional method measurement of sample~Ni(111) at normal AoI and  at wavelength 635\,nm. The solid lines represent the prediction of Yeh's 4$\times$4 transfer matrix numerical model using the MO parameters from Tab. \protect\ref{tab:RS111_fit_result}.}
\label{f:8dir_111RS_0AoI}
\end{figure}

In Fig. \ref{f:8dir_111RS_0AoI}, we show eight-directional method measurements of the sample~Ni(111) for normal AoI. In spite of a larger noise level in the first 120$^\circ$ range of the sample angle dependence, it can clearly be seen that the threefold angular dependencies of CMOKE, manifested in LMOKE+LCMOKE $\Phi_{M_L,M_L^3}$ and TCMOKE $\Phi_{M_T^3}$, are still present as they do depend on $A_p$ which is even in AoI and non-zero at normal incidence. On the other hand, the threefold angular dependencies of QMOKE $\Phi_{M_LM_T}$ and QMOKE $\Phi_{M_T^2-M_L^2}$ have vanished as they do depend on $B_p$, which is odd in AoI and thus zero at normal incidence. 

 The solid lines in Fig. \ref{f:8dir_111RS_0AoI} represent results calculated using Yeh's 4$\times$4 transfer matrix formalism, while all the parameters were fixed using values from Tab. \ref{tab:RS111_fit_result}, i.e. the solid lines in Fig. \ref{f:8dir_111RS_0AoI} are predictions of eight-directional method measurement at normal AoI, based on the numerical model fit to experimental data obtained from measurements at an AoI of 45$^\circ$ (Fig.~\ref{f:8dir_111RS}). We can see that this model prediction and the experiment are in decent agreement. The only exception is the Kerr rotation offset of QMOKE $\Phi_{M_LM_T}$, that is slightly different for the model and the experimental value. The numerical model prediction slightly overestimates the amplitude of the LMOKE+LCMOKE $\Phi_{M_L,M_L^3}$ and TCMOKE $\Phi_{M_T^3}$ angular dependencies. The very slight offset of the LMOKE+LCMOKE $\Phi_{M_L,M_L^3}$ contribution in the experimental data can be to some extent explained by not exactly normal AoI in the experiment. The offset of the TCMOKE $\Phi_{M_T^3}$ contribution is already known from Fig.~\ref{f:8dir_111RS} and being of an unknown origin at the moment.  

 %%%%%%%%%%%%%%%%%%%%%%%%%%%%%%%%%%%%%%%%%%%
%---Experimental Ni (001)
%%%%%%%%%%%%%%%%%%%%%%%%%%%%%%%%%%%%%%%%%%%
 \subsection{Eight-directional method measurement of the Ni(001) sample}
 
\begin{figure*}
\includegraphics{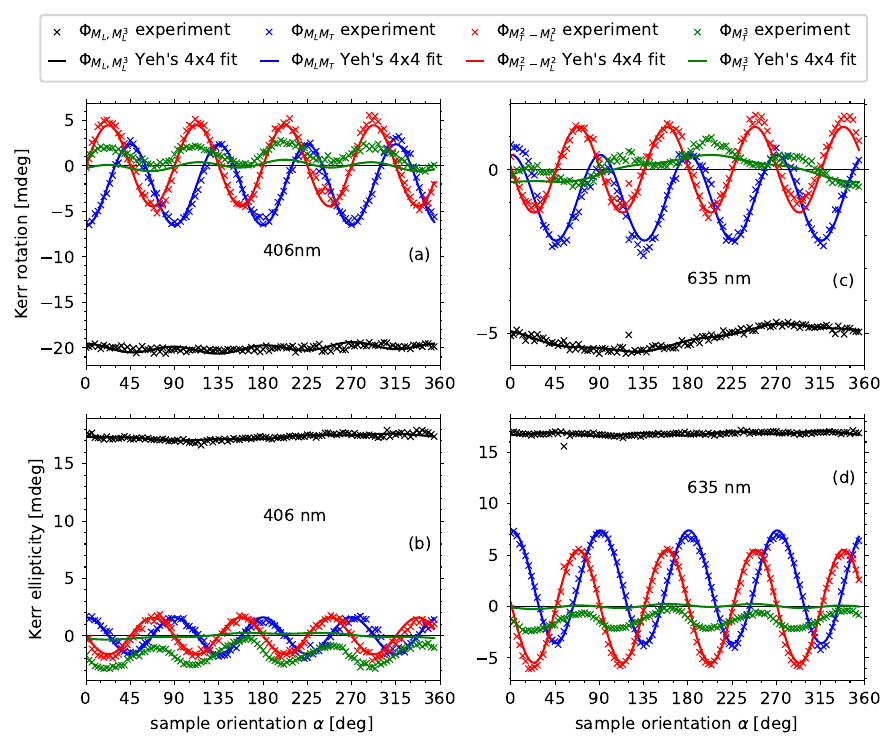}
\caption{Eight-directional method measurement of sample~Ni(001) with $p$-polarized light and at an AoI of 45$^\circ$. Graphs present the dependence on the in-plane sample orientation $\alpha$ of the (a) Kerr rotation and (b) Kerr ellipticity at a wavelength of 406\,nm as well as (c) Kerr rotation and (d) Kerr ellipticity at a wavelength of 635\,nm. Experimental data points are plotted as "x" marks, whereas solid lines represent the fit of the numerical model based on Yeh's 4$\times$4 transfer matrix formalism to the experimental data.}
\label{f:8dir_001RS}
\end{figure*}

In Fig. \ref{f:8dir_001RS}, we show the experimental results of the eight-directional method measurements for the sample~Ni(001), executed with $p$-polarized incident light and at an AoI of 45$^\circ$. For the wavelength of 406\,nm, individual contributions to the Kerr rotation and Kerr ellipticity, separated by eight-directional method, are plotted in Fig. \ref{f:8dir_001RS}(a) and Fig. \ref{f:8dir_001RS}(b), respectively. For the wavelength of 635\,nm, the contributions to Kerr rotation and Kerr ellipticity are plotted in Fig. \ref{f:8dir_001RS}(c) and Fig. \ref{f:8dir_001RS}(d), respectively. Analytical expressions for individual contributions given by Eq.~\eqref{eq:Kerr001_final} and Eqs. (\ref{eq:LCMOKE_measure}-\ref{eq:TCMOKE_measure}) are summarized in Tab.~\ref{tab:sample_Ni001_contributions}.

The QMOKE $\sim M_LM_T$ and QMOKE $\sim (M_T^2-M_L^2)$ anisotropy of (001)-oriented cubic crystal structures, manifested by four-fold angular dependencies, is well known and confirmed by various studies in literature \cite{Postava2002, Hamrle07a, Hamrle07b, Trudel10c, Wolf2011, Trudel11, Liang2015, Silber19}. With addition of the third order in magnetization contribution, Eq.~\eqref{eq:Kerr001_final} further predicts four-fold angular dependencies of CMOKE, observable within measurements of LCMOKE $\sim M_L^3$ and TCMOKE $\sim M_T^3$. In Fig. \ref{f:8dir_001RS}, such angular dependencies are experimentally observable only with the TCMOKE contribution, whereas no four-fold angular dependence is observable for the LMOKE+LCMOKE contribution. This is in contradiction with the description of Eq.~\eqref{eq:Kerr001_final}. The one-fold symmetry of LMOKE+LCMOKE and TCMOKE originates from VISMOKE as explained above and in Appendix~\ref{app:vismoke}.

 In Tab. \ref{tab:RS001_amplitude_offset} we present values of the offsets  and amplitudes $A_4$ of the four-fold angular dependencies of each MOKE contribution from Fig. \ref{f:8dir_001RS}. The values are provided by a fit of the simple model
 \begin{equation}
	\Phi_x(\alpha)=A_4\sin(4(\alpha-\alpha_4))+A_1\sin(\alpha-\alpha_1)+C
	\label{eq:simple_model_001}
\end{equation}
 to each MOKE contribution experimentally separated by the eight-directional method, where $x=\{(M_L,M_L^3), M_T^3, M_LM_T,(M_T^2-M_L^2)\}$. 
 
% For the LCMOKE$\sim M_L^3$ and TCMOKE$\sim M_T^3$  the values predicted by the numerical model with fixed $\Delta H$ are presented in the brackets in Tab. \ref{tab:RS001_amplitude_offset}. 

%TABLE AMPLITUDES AND OFFSETS FROM RS 001
\begin{center}
\begin{table}
\begin{tabular}{lrcrcrcrc}
			\hline
			\hline
			&&\multicolumn{7}{c}{Kerr rotation [mdeg]}\\
				\hline
				 && \multicolumn{3}{c}{406\,nm} && \multicolumn{3}{c}{635\,nm}   \\
				 \hline
			       && offset &&  \makecell{4-fold \\  amplitude}  && offset &&  \makecell{4-fold \\  amplitude} \\

			\hline\\
			 $M_L,M_L^3$	 && -20.00$\pm$0.02  && \makecell{0.08$\pm$0.03 \\  (0.38)} && -5.12$\pm$0.01 && \makecell{-0.02$\pm$0.02 \\  (0.05)} \\[4mm]
			 $M_T^3$	 && 1.16$\pm$0.02  && \makecell{1.08$\pm$0.03 \\  (0.38)} &&  0.13$\pm$0.01 && \makecell{0.30$\pm$0.01 \\  (0.05)} \\[4mm]
			 $M_LM_T$	 &&  -2.08$\pm$0.04  && -4.26$\pm$0.06 &&  -0.85$\pm$0.02 &&  1.35$\pm$0.03\\[1mm]
			 $M_T^2-M_L^2$	 &&  0.30$\pm$0.04   && 4.68$\pm$0.06 && 0.15$\pm$0.01 && -1.28$\pm$0.02 \\[1mm]
			 
			 			\hline
			&&\multicolumn{7}{c}{Kerr ellipticity [mdeg]}\\
				\hline
				 && \multicolumn{3}{c}{406\,nm} && \multicolumn{3}{c}{635\,nm}   \\
				 \hline
			       && offset &&  \makecell{4-fold \\  amplitude}  && offset &&  \makecell{4-fold \\  amplitude} \\

			\hline\\
			 $M_L,M_L^3$	  && 17.30$\pm$0.01  && \makecell{-0.10$\pm$0.02 \\  (-0.07)} && 16.82$\pm$0.01 && \makecell{0.07$\pm$0.02 \\  (-0.16)} \\[4mm]
			 $M_T^3$	 && -1.66$\pm$0.02  && \makecell{-0.95$\pm$0.03 \\  (-0.07)} &&  -0.13$\pm$0.02 && \makecell{-0.86$\pm$0.02\\  (-0.16)} \\[4mm]
			 $M_LM_T$	 &&  -0.02$\pm$0.02  && 1.53$\pm$0.03 &&  1.91$\pm$0.03 &&  5.36$\pm$0.04\\[1mm]
			 $M_T^2-M_L^2$	 &&  -0.13$\pm$0.02   && -1.68$\pm$0.03 && -0.23$\pm$0.03 && -5.65$\pm$0.04 \\[1mm]
		\hline
		\hline
\end{tabular}
\caption{Values of 4-fold amplitudes and offsets from eight-directional measurement of sample~Ni(001) presented in Fig. \protect\ref{f:8dir_001RS}. The numbers in brackets are the values of the four-fold angular dependencies amplitudes $A4$ predicted by the numerical model with fixed $\Delta H$ (see description in the main text).}
\label{tab:RS001_amplitude_offset}
\end{table}
\end{center}

The solid lines in Fig. \ref{f:8dir_001RS} present Yeh's 4$\times$4 transfer matrix numerical model fit to the experimental data. We set $K, G_s$ and $2G_{44}$ (together with vicinal parameters $\varepsilon_S$ and $\alpha_1)$ as free parameters. Because there is a difference in the amplitude of LCMOKE and TCMOKE angular dependencies, which the model (analytical as well as numerical) cannot describe, we decided to fix $\Delta H$ to its value from Tab. \ref{tab:RS111_fit_result}. See that the predicted amplitudes, resulting from the values in Tab. \ref{tab:RS111_fit_result}, favour the LCMOKE$\sim M_L^3$ experimental measurements over TCMOKE$\sim M_T^3$ experimental findings. Note that the values of MO parameters of the same material can differ more or less when gathered from different samples \cite{Silber19} as it is the case here (compare values in Tab. \ref{tab:RS111_fit_result} and \ref{tab:RS001_fit_result}). Thus, fixing $\Delta H$ to the value obtained from sample~Ni(111) give us only a rough estimate of the CMOKE angular dependence in the sample~Ni(001), but the order of magnitude should hold. Furthermore, it also allows us to show how the same values of the cubic MO parameters manifest in (001)- compared to (111)-oriented cubic crystal structures. It is shown that in an (001)-oriented cubic crystal structure, the CMOKE effect is less pronounced, due to the prefactor $B_{p}$ as will be discussed later in the text. For the LCMOKE$\sim M_L^3$ and TCMOKE$\sim M_T^3$  the values of the four-fold angular dependencies amplitudes $A_4$ predicted by the numerical model with fixed $\Delta H$ are presented in the brackets in Tab. \ref{tab:RS001_amplitude_offset}. 
% while the QMOKE is weighted by the larger prefactor $A_{s/p}$. The situation is thus vice versa compared to (111)-oriented crystal structures. 
 
\begin{center}
\begin{table*}
\begin{tabular}{lrcrcrcrc}
			\hline
			\hline
			MO parameter      && 406 nm && 635 nm  \\
			\hline \\
			 $K$	 &&  -0.0753$\pm$0.0003\,-(0.0505$\pm$0.0003)i  && -0.2756$\pm$0.0008\,-(0.0370$\pm$0.0008)i  \\[1mm]
			 $G_s$	 && 0.0017$\pm$0.0001\,-(0.0022$\pm$0.0001)i  && -0.0069$\pm$0.0001\,-(0.0079$\pm$0.0001)i  \\[1mm]
			 $2G_{44}$	 &&  -0.0037$\pm$0.0001\,+(0.0020$\pm$0.0001)i  && 0.0035$\pm$0.0001\,+(0.0095$\pm$0.0001)i\\[1mm]
			 $\Delta G$	 &&  0.0054$\pm$0.0002\,-(0.0042$\pm$0.0002)i  && -0.0104$\pm$0.0002\,-(0.0174$\pm$0.0002)i\\[1mm]
			 $\Delta H$ (fixed)	 &&  0.0023+0.0046i   && 0.0100 +0.0018i \\[1mm]
			 \hline
			 vicinal parameters && && \\
			 \hline
			 $\varepsilon_S$	 &&  -0.0130$\pm$0.0068\,+(0.0296$\pm$0.0068)i  &&  -0.0299$\pm$0.0080\,+(0.0557$\pm$0.0080)i \\[1mm]
			 $\alpha_1$  && -22$^\circ$$\pm$12$^\circ$ && -20$^\circ$$\pm$7$^\circ$ \\
%		$\Delta\alpha$ && 0$^\circ$ && 0 $^\circ$ \\
		\hline
		\hline
\end{tabular}
\caption{Values of MO parameters for Ni layer of sample~Ni(001), provided by the fit of Yeh's 4$\times$4 transfer matrix numerical model to the experimental data as is shown in Fig. \protect\ref{f:8dir_001RS}. The value of $\Delta G = G_s-2G_{44}$ is shown for comparison with $\Delta H$, where $\Delta H$ is fixed to the value from Tab. \ref{tab:RS111_fit_result} . Value of $\varepsilon_S$ and $\alpha_1$ are related to VISMOKE as described in Appendix \ref{app:vismoke}.}
\label{tab:RS001_fit_result}
\end{table*}
\end{center}

%Finally, in Fig. \ref{f:8dir_001RS_0aoi} the eight-directional method measurement of the sample~Ni(001) at normal AoI and for wavelength of 635\,nm is shown for both Kerr rotation and Kerr ellipticity. Only QMOKE$\sim M_LM_T$ and QMOKE$\sim M_T^2-M_L^2$ persisted with normal AoI, as predicted by the theory. The solid lines in the graphs are again the normal AoI predictions of the numerical model that use parameters from Tab. \ref{tab:RS001_fit_result}. The model under-/overestimate the amplitudes of the QMOKE angular dependence for Kerr rotation/ellipticity, respectively. The likely reason is that the MO parameters are not found as precisely with sample~Ni(001) as with the sample~Ni(111).
%\begin{figure}
%\includegraphics{fig_source/8dir_RS001_AoI0.pdf}
%\caption{Eight-directional method measurement of sample~Ni(001), at normal AoI anda wavelength of 635\,nm. Experimental data are plotted as "x" marks. The full lines is prediction based on Yeh's 4$\times$4 transfer matrix numerical model, using values from Tab.~\protect\ref{tab:RS001_fit_result}. }
%\label{f:8dir_001RS_0aoi}
%\end{figure}

\section{Discussion}

\subsection{Exclusion of parasitic angular-denpendent effects}

 As presented above, the theory of CMOKE describes the observed anisotropies of the sample~Ni(111) very well, but struggles with the sample~Ni(001) for which the TCMOKE angular dependencies possess a different amplitude than the LCMOKE contribution, which cannot be explained by the CMOKE contribution. Thus, one can question if there is another effect causing those LCMOKE and TCMOKE angular dependencies with sample~Ni(111).

First, the effect of in-plane magnetocrystalline anisotropy could provide similar angular dependencies as observed with LCMOKE $\sim M_L^3$ and TCMOKE $\sim M_T^3$ if for some sample orientations the $\bm{M}$ was not fully saturated by the external magnetic field in longitudinal, or transversal direction, while for another sample orientation $\alpha$ it was. Such an in-plane  magnetocrystalline anisotropy could have three-fold symmetry for the sample~Ni(111), exactly as is the symmetry of the QMOKE and CMOKE contributions. However, note that if this were the case, the angular dependencies of LCMOKE and TCMOKE should vanish for normal AoI, since LMOKE$\sim B_{s/p}$. As is shown in Fig. \ref{f:8dir_111RS_0AoI}, the three-fold angular dependencies of LCMOKE and TCMOKE are well pronounced at normal AoI, which corresponds to the CMOKE theory in which this angular dependence is proportional to the optical factor $A_{s/p}$. In addition, as discussed in previous work \cite{Gaerner24}, the angular dependence of magnetic remanence and coercivity show a two-fold symmetry and not a three-fold one, supporting the claim that the magnetocrystalline anisotropy cannot cause the angular dependence explained as CMOKE here.

Second, our equations and Yeh's 4$\times$4 transfer matrix model assume only in-plane magnetization, while the out-of-plane component of $\bm{M}$ is set equal to zero. As discussed in Sec.~\ref{theory_moke}, for (001)-oriented cubic crystal structures no new anisotropies are induced when $M_P$ is being considered. For (111)-oriented cubic crystal structures, there are new QMOKE anisotropic terms $M_LM_P$ and $M_TM_P$ (being discussed at the end of Sec.~\ref{sec:Kerr111_final}), which possess the same dependence on the sample orientation $\alpha$ as the CMOKE contribution does. Nevertheless, if the $M_P$ component of $\bm{M}$ were present alongside with the $M_L$ component, while a longitudinal external magnetic field is applied, there would be a considerable PMOKE $\sim M_P$ contribution in form of an offset. This is not observed within our measurement at normal AoI presented in Fig. \ref{f:8dir_111RS_0AoI}. Furthermore, the sign of potential QMOKE anisotropic terms $M_LM_P$ and $M_TM_P$ would be sensitive to the change of polarization of the incident light, whereas the CMOKE contribution is not. In Appendix \ref{app:Ni111_polarizations} we show a comparison of eight-directional method measurements of sample Ni(111) measured with $s$- and $p$-polarized light at wavelengths of 488nm and 670nm. Note that the sign of the amplitude of LCMOKE $\sim M_L^3$ and TCMOKE $\sim M_T^3$ is not affected by the change of the polarization, just as predicted by the theory of CMOKE (see Tab. \ref{tab:sample_Ni111_contributions}). Thus this excludes the possibility that the observed angular dependencies are of QMOKE $\sim M_LM_P$ and QMOKE $\sim M_TM_P$ origin.

Third, higher-order-in-magnetization contributions of VISMOKE might mimic the CMOKE angular dependence. However, this also cannot be the source of the three-fold angular dependencies of LCMOKE $\sim M_L^3$ and TCMOKE $\sim M_T^3$ in the sample~Ni(111). As discussed in appendix \ref{app:vismoke}, the symmetry of such higher-order VISMOKE would be different. Moreover, the effect of such a contribution is  already covered within the numerical model using Yeh's 4$\times$4 transfer matrix formalism, and the numerical model with VISMOKE effect but without the parameter $\Delta H$ is not able to describe the data of sample~Ni(111), as we have tried unsuccessfully.

We therefore conclude that the presented three-fold angular dependencies of LCMOKE $\sim M_L^3$ and TCMOKE $\sim M_T^3$ with the sample~Ni(111) shown in Fig. \ref{f:8dir_111RS} are stemming from CMOKE described by a MO parameter $\Delta H + K\Delta G /\varepsilon_d$, in which $\Delta H$ plays the major role \cite{Gaerner24}. The reason why no four-fold angular dependence is observed in the measurement of LMOKE+LCMOKE $\sim M_L,M_L^3$ with the sample Ni(001) may be due to the $B_{s/p}$ optical weighting factors that are present with CMOKE angular dependencies for the (001)-oriented cubic crystal structure compared to the $A_{s/p}$ optical weighting factors that are present with CMOKE angular dependencies for (111)-oriented cubic crystal structures as shown in Tab. \ref{tab:sample_Ni001_contributions} and \ref{tab:sample_Ni111_contributions}. 

\subsection{The role of optical weighting factors $A_{s/p}$ and $B_{s/p}$ for different CMOKE contributions}

\begin{figure*}
\includegraphics{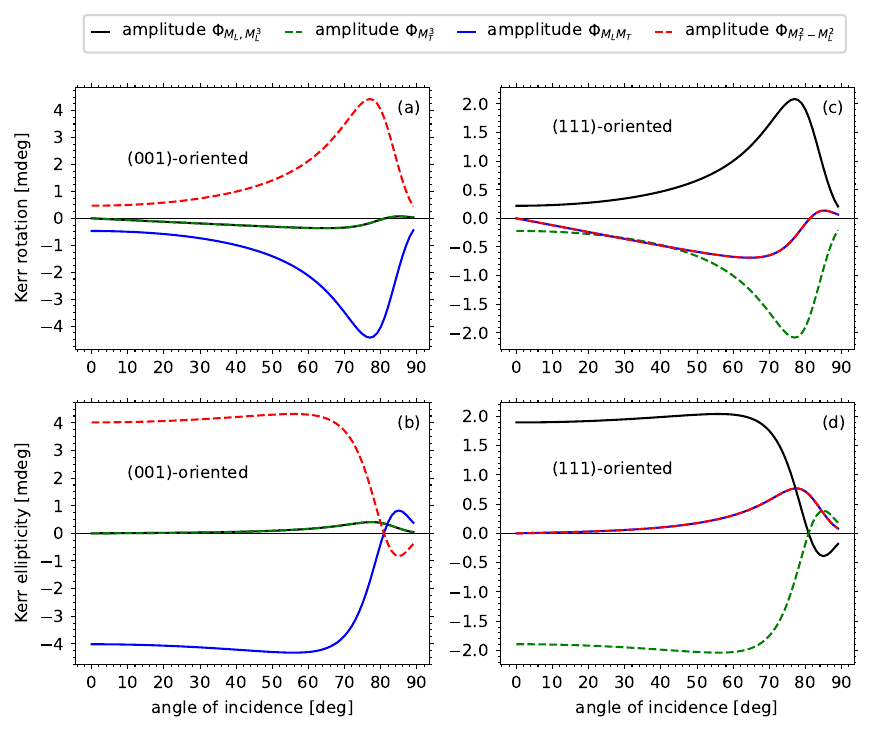}
\caption{(a),(c) Kerr rotation and (b),(d) Kerr ellipticity of the amplitude of (a),(b) the four-fold angular dependence for (001)-oriented cubic crystal structure and (c),(d) the three-fold angular dependence for (111)-oriented cubic crystal structure depending on the AoI. The data were simulated using the numerical model based on Yeh's $4\times 4$ matrix formalism, for which structural and optical parameters of sample Ni(111) were used, but  MO parameters were chosen to be $K=0+0i$, $\Delta G=0.01+0.01i$ (with $G_s=0+0i$ and $2G_{44}=0.01+0.01i$) and $\Delta H=0.01+0.01i$ (with $H_{123}=0.01+0.01i$ and $H_{125}=0+0i$). Vicinal structural parameter $\varepsilon_S=0$ here. The simulation is done for a wavelength of 635\,nm.} 
\label{f:Ni_MOparset_AoI}
\end{figure*}

To better demonstrate this effect of optical weighting factors,  we performed the simulations (based on Yeh's $4\times 4$ matrix formalism) of the eight-directional method and its dependence on AoI for (001)- and (111)-oriented cubic crystal structures. For both crystallographic orientations we use the exact same input parameters. The MO parameters were chosen to be $K=0+0i$, $G_s=0.01+0.01i$, $2G_{44}=0+0i$, $H_{123}=0.01+0.01i$ and $3H_{125}=0+0i$, which results in $\Delta G=\Delta H= 0.01+0.01i$.  With such MO parameters, we simulate only QMOKE angular dependencies that stem from $\Delta G$ and CMOKE angular dependencies that stem from $\Delta H$, with suppressed contribution $K\Delta G/\varepsilon_d$ in the case of CMOKE with (111)- oriented cubic crystal structure. Structural and optical parameters were taken from sample Ni(111), while VISMOKE was not considered at all in this simulation and therefore $\varepsilon_S=0$. The simulations were performed at a wavelength of 635nm and for $p$-polarized light. This way we can clearly compare the effect of the optical weighting factors $A_{p}$ and $B_{p}$, how the amplitude of the CMOKE changes between (001)- and (111)-oriented cubic crystal structures and how the value of the CMOKE amplitude compares to the value of the QMOKE amplitude with each crystallographic orientation. 

In Fig. \ref{f:Ni_MOparset_AoI}, we show the resulting values of the amplitudes of the QMOKE and CMOKE angular dependencies and their dependence on the AoI for (001)- and (111)-oriented cubic crystal structures. In Figs. \ref{f:Ni_MOparset_AoI}(a) and (b), Kerr rotation and Kerr ellipticity are shown, respectively, for the four-fold angular dependence amplitude of each contribution separated by the eight-directional method with (001)-oriented cubic crystal structure.  In Fig. \ref{f:Ni_MOparset_AoI}(c) and (d), Kerr rotation and Kerr ellipticity are shown, respectively, for three-fold angular dependence amplitudes of each contribution separated by the eight-directional method with (111)- oriented cubic crystal structure. 

For the (001)-oriented cubic crystal structure, the angular dependence amplitude $A_4$ of CMOKE is about one order of magnitude smaller than that of QMOKE. The only exception is the Kerr ellipticity for an AoI around $80^\circ$, for which CMOKE and QMOKE angular dependence amplitudes can be comparable due to a decrease in the value of the QMOKE amplitude as seen in Fig. \ref{f:Ni_MOparset_AoI}(b). Considering that in real samples $\Delta H$ is likely to be smaller than $\Delta G$ (as a rule of thumb MO parameters tend to become weaker with higher order in magnetization), and that even the detection of QMOKE angular dependence is not a trivial MOKE measurement in itself, it implies that the detection of CMOKE with (001)-oriented cubic structures will be a challenging task and can easily be lost or overlooked in the signal noise. This may explain why a four-fold angular dependence of LMOKE+LCMOKE has not been observed so far in (001)-oriented cubic crystal structures.  Although a slight fourfold angular dependence of the LMOKE signal has been reported in measurements on (001)-oriented cubic crystals \cite{Trudel10c, Trudel_10d, Trudel_10e, Trudel11, Wolf2011} the authors did not offer a clear explanation for the observed behaviour and we can rule out CMOKE as the origin here, since all data were taken at normal incidence. Measurements at grazing AoI would be a suitable approach to detect CMOKE in (001)-oriented cubic crystal structures, which is being in the scope of the future research. 
%To our knowledge, there is no report in the literature of four-fold angular dependence of the LMOKE contribution in the eight-directional method measurement of (001)-oriented cubic crystal samples.The only exception is published here \cite{Wolf2011}, but no explanation is given and, as it is the case of LMOKE measurement with normal AoI, the CMOKE angular dependence should go extinct here anyway. 

With (111)-oriented cubic crystal structures the situation is quite different. It can be seen that the amplitude of the CMOKE angular dependence is even more pronounced than the QMOKE amplitude. For the Kerr rotation in Fig. \ref{f:Ni_MOparset_AoI}(c) the CMOKE and QMOKE amplitudes are of comparable size between 20$^\circ$ to 50$^\circ$, but elsewhere the CMOKE amplitude is stronger. For Kerr ellipticity, the CMOKE amplitude is dominant up to an AoI of about 75$^\circ$, and for more grazing AoIs, the CMOKE amplitude decreases and is comparable to the QMOKE amplitude. This suggests that in (111)-oriented cubic crystal structures, the CMOKE amplitude is at least equal to, but mostly stronger than that of QMOKE. For example in Fig. \ref{f:8dir_111RS}(d), the QMOKE angular dependencies are barely observable but the CMOKE angular dependencies are clearly pronounced. Still, the dependence on the photon wavelength plays an important role as well and can change the described behaviour. It is therefore absolutely necessary to explore CMOKE spectroscopy in the future and compare CMOKE spectra to QMOKE spectra in order to find energies of dominating CMOKE vs. dominating QMOKE for various crystal orientations. Finally, note that CMOKE has advantages over QMOKE for (111)-oriented cubic crystal structures as the CMOKE has a finite angular dependence amplitude at the normal AoI, which can be leveraged in various applications, as discussed in the introduction of this paper. 

\subsection{TCMOKE angular dependence with sample~Ni(001)}

Because we are not able to explain the strong amplitude of the TCMOKE angular dependence with sample~Ni(001), we tend to think that it is not solely due to the CMOKE effect. The reason is that the amplitudes of LCMOKE $\sim M_L^3$ and TCMOKE $\sim M_T^3$ angular dependence do not match each other and to describe the TCMOKE angular dependence for the sample~Ni(001), a significantly larger $\Delta H$ would be required than found for the sample~Ni(111). Finally, it should be noted that the angular dependence of the TCMOKE with sample~Ni(001) does not appear to be purely sinusoidal (see Fig. \ref{f:8dir_001RS}(a) and (d)). Furthermore, the offset of the TCMOKE contribution presented in both samples cannot be explained by theory. As the measurement of MOKE with transversal directions, i.e. Eq.~\eqref{eq:TCMOKE_measure}, is not usually presented in the literature when using the eight-directional method, we are unable to compare our experimental observations with the other published results. The possible source of the observed effect can be different strain within the Ni(001) layer grown on MgO(001) substrate compared to the Ni(111) layer grown on MgO(111) substrate. Therefore, further experimental analysis of the angular dependence of TCMOKE is needed to determine its additional origin.

\section{Conclusion}
We provided a detailed description of the cubic-in-magnetization contributions to the MOKE for (001)- and (111)-oriented cubic crystal structures.  The anisotropy of CMOKE, predicted by the developed equations, is compared to  experimental measurements of two fcc Ni samples with (001)- and (111)-oriented cubic crystal structure orientation. The experimental data are also described phenomenologically using Yeh's 4$\times$4 transfer matrix formalism. The MO parameters $H_{123}$ and $H_{125}$ are used to describe CMOKE contributions in addition to the MO parameter $K$ describing LinMOKE and the MO parameters $G_s$ and $2G_{44}$ describing QMOKE. The parameter $\Delta H = H_{123}-3H_{125}$ is then used to describe the anisotropy of the CMOKE effect. Only the offsets of certain contributions as well as the amplitude of the TCMOKE angular dependence in the sample~Ni(001) could not be described well. Otherwise all the anisotropies and AoI dependencies, as well as dependencies on the polarization of incident light, are well predicted by the model of CMOKE when compared to the experimental observations of the Ni samples. 

By comparing the equations for the (001)- and (111)-oriented cubic crystal structures, and using the value of $\Delta H$ found from the experiment, we conclude that CMOKE is being well pronounced for the (111)-oriented cubic crystal structures, but is suppressed for the (001)-oriented cubic crystal structures, where the angular dependence  of QMOKE dominates. Furthermore, for the (001)-oriented cubic crystal structures, the CMOKE angular dependence disappears with normal AoI but the QMOKE angular dependence persists.  However, for (111)-oriented cubic crystal structures the situation is reversed. Here, the CMOKE amplitudes of the angular dependencies are more pronounced than the QMOKE angular dependencies and are also present at normal AoI, whereas the QMOKE amplitudes vanish. Thus, it is conclusive that CMOKE has been recently observed in (111)-oriented samples \cite{Gaerner24, pan25}, while a clear experimental observation of CMOKE in (001)-oriented cubic crystal structures is still pending.

%%%%%%%%%%%%%%%%%%%%%%%%%%%%%%%%%%%%%%%%%
%---ACKNOWLEDGEMENT
%%%%%%%%%%%%%%%%%%%%%%%%%%%%%%%%%%%%%%%%%%%
\begin{acknowledgments}
We acknowledge the project  “Materials and Technologies for Sustainable Development,” (CZ02.01.01/00/22\_008/0004631) funded by the European Union and the state budget of the Czech Republic within the framework of the Jan Amos Komensk\'{y} Operational Program, project "Advanced Materials for Energy and Environmental Technologies" (CZ.02.01.01/00/23\_021/0008592) and project "REFRESH" (CZ.10.03.01/00/22\_003/0000048). Further support was provided by GACR (25-15775S); Ministry of Education, Youth and Sports of the Czech Republic (SP2025/090).

 We thank G. Reiss for making available laboratory equipment.
\end{acknowledgments}

\section*{Data availability}
All data supporting the findings of this study are available for download at the Zenodo repository \cite{datarepo}.

%%%%%%%%%%%%%%%%%%%%%%%%%%%%%%%%%%%%%%%%%%%
%---APPENDIX
%%%%%%%%%%%%%%%%%%%%%%%%%%%%%%%%%%%%%%%%%%%
\appendix
\renewcommand\thefigure{\thesection.\arabic{figure}} 

\section{Sign conventions}
\label{app:conventions}
\setcounter{figure}{0}  

\begin{figure}
\includegraphics{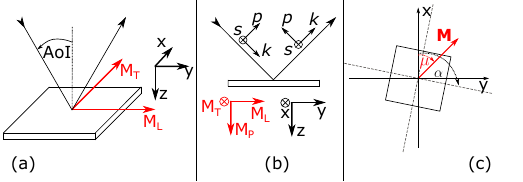}
\caption{(a) The plane of incidence and the surface of the sample define the Cartesian coordinate system $\hat{x}$, $\hat{y}$, $\hat{z}$, where $\hat{x}$ is perpendicular to the plane of incidence and parallel to the surface of the sample. (b) Definition of the $\hat{s}$, $\hat{p}$, $\hat{k}$ Cartesian system  for the incident and reflected beam and definition of the directions of the  normalized magnetization components $M_T \parallel \hat{x}$, $M_L\parallel \hat{y}$ and $M_P\parallel \hat{z}$. (c) Visualisation of the positive in-plane rotation of the sample and the magnetization within the $\hat{x}$, $\hat{y}$, $\hat{z}$ coordinate system, described by the angle $\alpha$ and $\mu$, respectively. The figure is taken from Ref. \cite{Silber19} and slightly modified.}
\label{f:xyz_sys}
\end{figure}

  We use the negative time convention as $\bm{E}{(\bm{r},t)}=\bm{E}(\bm{r}) e^{-i\omega t}$, giving the permittivity in the form $\varepsilon=\Re(\varepsilon) +i\Im(\varepsilon)$, where the imaginary part of the complex permittivity $\Im(\varepsilon) >0$. The rotations of the crystallographic structure, sample and magnetization take place in the right-handed  $\hat{x}$, $\hat{y}$, $\hat{z}$ system  and the incident and reflected light beams are described in the $\hat{s}$, $\hat{p}$, $\hat{k}$ system as is shown in Fig.~\ref{f:xyz_sys}. The positive rotation of the sample and magnetization is clockwise, when looking at the top surface of the sample (i.e. the vector pointing in the $\hat{x}$  direction rotates towards the $\hat{y}$). The reflection from the sample is described in Jones formalism by the reflection matrix
 \begin{equation}
 	\bm{R}=
 	\begin{bmatrix}
 	r_{ss}&r_{sp}\\
 	r_{ps}&r_{pp}	
 	\end{bmatrix}.
 \end{equation}
	
Cartesian vector rotation about $\hat{z}$ axis by angle $\alpha$ is expressed as
\begin{equation}
	\bm{a}^{\hat{z}}=
		\begin{bmatrix}
		\cos{\alpha} & -\sin{\alpha} &0\\
		\sin{\alpha} & \cos{\alpha} &0\\
		0&0&1\\
	\end{bmatrix},
	\label{eq:a_rot_z}
\end{equation}
respectively.

For a (001)-oriented cubic crystal structure, the reference orientation (i.e. sample orientation $\alpha$ = 0) corresponds to the [100] direction being parallel to the $\hat{x}$-axis of our coordinate system and the [001] direction being parallel to the $\hat{z}$-axis of our coordinate system. For a (111)-oriented cubic crystal structure, the reference orientation (i.e. sample orientation $\alpha$ = 0) corresponds to the [-211] direction being parallel to the $\hat{x}$-axis of our coordinate system and the [111] direction being parallel to the $\hat{z}$-axis of our coordinate system. To transform between those two reference orientations, the transformation matrix of Eq.~\ref{eq:a_rot_to_111} is used.

For a more detailed description of the sign conventions, see Ref. \cite{Silber19}.

\section{Derivation of the $\bf{\textit{H}}$ tensor}
\label{app:H_tensor}

The derivation steps of the MO tensor $\bm{H}$ were inspired by the derivation of the MO tensors $\bm{K}$ and $\bm{G}$ in Refs. \cite{Visnovsky1986, Visnovsky06}. Upon consideration of the Onsager relation $\varepsilon_{ij}(\omega,\bm{M})=\varepsilon_{ji}(\omega,-\bm{M})$, where $\omega$ is the frequency of the electromagnetic wave, we can write the following rules for the elements of the MO tensor $\bm{H}$.

\begin{equation}
  H_{iiklm}=0,
\end{equation}
and is invariant to permutations $k\leftrightarrow l \leftrightarrow m$ and antisymmetric with $i\leftrightarrow j$, which results in
\begin{equation}
\begin{split}	
  H_{ijklm}=H_{ijkml}=H_{ijlkm}&{=}H_{ijlmk}=H_{ijmkl}=H_{ijmlk}\\ 
  =-H_{jiklm}=-H_{jikml}&{=}-H_{jilkm}=-H_{jilmk}\\
  =-H_{jimkl}&{=}-H_{jimlk}.
 \end{split}
 \end{equation}

\noindent
Further, the following symmetry operations apply: rotation around twofold axes $\hat{x}$, $\hat{y}$ and $\hat{z}$, being expressed as
\begin{equation}
\begin{split}
C_{2x} =
\begin{bmatrix}
1&0&0\\
0&-1&0\\
0&0&-1
\end{bmatrix}&{,}\,\,
C_{2y} =
\begin{bmatrix}
-1&0&0\\
0&1&0\\
0&0&-1
\end{bmatrix},\\
 \text{and}\quad
C_{2z} &{=}
\begin{bmatrix}
-1&0&0\\
0&-1&0\\
0&0&1
\end{bmatrix},
\end{split}
\end{equation}
respectively. Moreover, we can apply a rotation around the threefold axis $C_3$ (around [111] direction) and a rotation around a twofold axis $C_{2a}$ (around [011] direction).
\begin{equation}
C_{3} =
\begin{bmatrix}
0&0&1\\
1&0&0\\
0&1&0
\end{bmatrix},\quad
C_{2a} =
\begin{bmatrix}
0&1&0\\
1&0&0\\
0&0&-1
\end{bmatrix}.
\end{equation}

All the symmetry transformations mentioned above, applied to the tensor $\bm{H}$ using Eq.~\eqref{eq:h_rot}, must satisfy $\bm{H}$=$\bm{H'}$.

\section{Structural characterization of sample Ni(001)}
\label{app:struct_char_001}

To characterize the sample~Ni(001), XRR and XRD measurements have been performed using  Phillips X$'$pert Pro MPD PW3040-60 with a Cu $K_\alpha$ source. 

\begin{figure}
\includegraphics{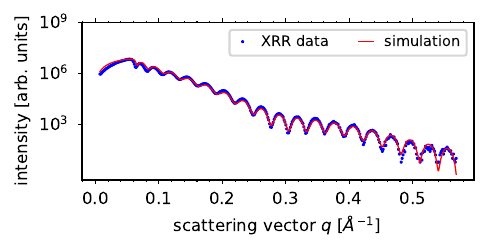}
\caption{XRR measurement and simulations of the sample~Ni(001): MgO(001)// Ni(18.8\,nm)/ Pt(2.2\,nm).}
\label{f:XRR_001}
\end{figure}

We used the open-source program GenX \cite{Bjorck2007} based on the Parratt algorithm \cite{Parratt1954} to analyze the XRR curves. The measured XRR curve and its simulations are displayed in Fig. \ref{f:XRR_001}. The fit for the XRR data in Fig. \ref{f:XRR_001} provides: substrate MgO roughness of 0.2\,nm, thickness of the Ni layer 18.8\,nm, roughness of the Ni layer 0.6\,nm, thickness of the Pt capping layer 2.2\,nm and roughness of the Pt capping layer of 0.6\,nm. For the XRR measurements and results of the sample Ni(111), please see Ref. \cite{Gaerner24}.

\begin{figure}
\includegraphics{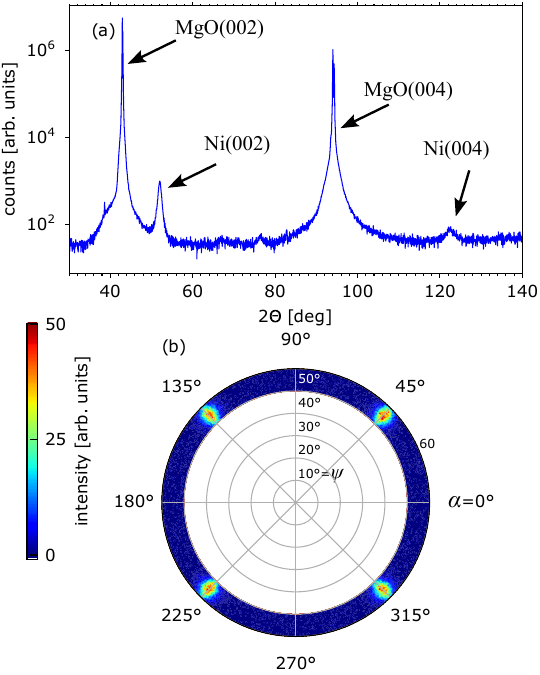}
\caption{(a) Specular XRD scan $\Theta$-$2\Theta$ of the sample~Ni(001). (b) An off-specular XRD scan (Euler cradle texture map) for $2\Theta=44.479$ (corresponding to Ni$\{$111$\}$ diffraction peak),  and $\Psi=\langle 50,60 \rangle$. }
\label{f:XRD_001}
\end{figure}

Further, in Fig. \ref{f:XRD_001}(a) a specular $\Theta$-$2\Theta$ scan of the sample~Ni(001) is presented. The profound peak around $2\Theta=52.0^\circ$ belongs to the Ni(002) diffraction peak and indicates good crystallinity of the sample. In Fig. \ref{f:XRD_001}(b), an off-specular XRD Euler cradle texture map is displayed. With $2\Theta$ set to $44.479^\circ$ and a tilt of $\Psi=\langle 50, 60 \rangle$, four Ni$\{111\}$ peaks are visible, showing the in-plane unified growth direction. For the XRD analysis of sample~Ni(111), see Ref. \cite{Gaerner24}.

\section{Ellipsometry}
\label{app:ellipsometry}

\begin{figure}
\includegraphics{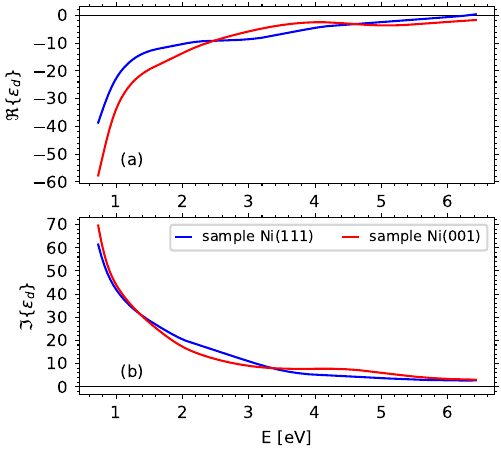}
\caption{Permittivity spectra of (a) real and (b) imaginary part of $\varepsilon_d$ for sample~Ni(111) and sample~Ni(001).}
\label{f:elipsometrie}
\end{figure}

Both samples have been measured using a M\"{u}ller matrix spectral ellipsometer Woolam RC2. When processing the measured data (using the Woolam RC2  complementary software CompleteEASE \cite{J.A.WoollamCo.2008}) the thicknesses of all the layers in both samples were fixed in the model, using the values provided by XRR. The permittivity spectra of the MgO substrates and of the Pt cap were taken from the CompleteEASE library \cite{Palik}. Permittivity spectra of the Si/SiO$_x$ cap was determined from a separate MgO/Si/SiO$_x$ sample. Then, the imaginary part of the permittivity spectra of the Ni layer in both samples were described by B-spline \cite{Johs2008}, while complementary spectra of the real part were determined through Kramers-Kronig relations. The resulting spectra for the real and imaginary part of the permittivity $\varepsilon_d$ are shown in Figs. \ref{f:elipsometrie}(a) and \ref{f:elipsometrie}(b), respectively. In Tab. \ref{tab:permitivity_Ni001} we summarize the values of permittivity of each layer used in our model to describe sample Ni(001) and in Tab. \ref{tab:permitivity_Ni111} we summarize the values of premittivity of each layer used in our model to describe sample Ni(111).

\begin{center}
\begin{table}
\begin{tabular}{lrcrcrcrc}
			\hline
			\hline
			layer      && 406 nm && 635 nm  \\

			\hline
			 MgO	 && 3.10+0.00i  && 3.01+0.00i  \\[1mm]
			 Ni	 && -5.48+8.81i  && -14.21+18.10i  \\[1mm]
			 Pt	 &&  -5.29+10.00i  &&  -11.87+19.49i	 \\[1mm]
		\hline
		\hline
\end{tabular}
\caption{The values of permittivity for sample Ni(001) at wavelength of 406\,nm and wavelength of 635\,nm.}
\label{tab:permitivity_Ni001}
\end{table}
\end{center}

\begin{center}
\begin{table}
\begin{tabular}{lrcrcrcrc}
			\hline
			\hline
			layer      && 406 nm && 635 nm  \\

			\hline
			 MgO	 && 3.10+0.00i  && 3.01+0.00i  \\[1mm]
			 Ni	 && -8.48+10.50i  && -10.59+21.12i \\[1mm]
			 Si	 && 10.91+8.55i  && 10.74+1.36i\\[1mm]
			 SiO$_x$	 &&  3.04+0.00i  && 3.01+0.00i\\[1mm]
		\hline
		\hline
\end{tabular}
\caption{The values of permittivity for sample Ni(111) at wavelength of 406\,nm and wavelength of 635\,nm.}
\label{tab:permitivity_Ni111}
\end{table}
\end{center}

\section{Vicinal interface sensitive MOKE}
\label{app:vismoke}

VISMOKE originates from interference between optostructural and MO perturbation, when a ferromagnetic film is deposited on the vicinal surface of the substrate \cite{Hamrle03}. The optostructural contribution can be effectively described by
\begin{equation}
\bm{\varepsilon}_S=
\begin{bmatrix}
0 &0&\varepsilon_s \cos(\alpha_1)\\
0& 0 &\varepsilon_s\sin(\alpha_1)\\
\varepsilon_s \cos(\alpha_1)&\varepsilon_s \sin(\alpha_1) &0\\	
\end{bmatrix},
\end{equation}
with $\alpha_1$ as the angle between $\hat{S}$ and $\hat{x}$, while $\hat{S}$ is defined by the intersection of a mirror plane of a stepped surface (which is perpendicular to the edges of the steps) and the surface plane of the sample at a reference orientation as defined in Appendix \ref{app:conventions}. Thus, $\alpha_1$ defines the in-plane orientation of the vicinal steps when the sample is at its reference orientation. The permittivity of the ferromagnetic layer is then  described as $\bm{\varepsilon}=\bm{\varepsilon}^{(0)}+\bm{\varepsilon}_S+\bm{\varepsilon}_M $, while $\bm{\varepsilon}_M$ is the MO change of the permittivity described in detail in Sec.~\ref{theory_moke}. The one-fold angular dependence observed in our experimental data of the eight-directional measurement is then linked to the product $\varepsilon_{yz/zy} \varepsilon_{zx/xz}$, which will provide terms as $\varepsilon_s\sin(\alpha_1)K\sin(\alpha)M_L$ and $\varepsilon_s\cos(\alpha_1)K\cos(\alpha)M_T$. Any product of $\varepsilon_s$ and higher order MO parameters is also possible and is taken care of by Yeh's 4$\times$4 matrix numerical model, while $\varepsilon_S$ and $\alpha_1$ are set as a free parameters. Note that the symmetry of such higher-order VISMOKE will not be the same as the symmetry of QMOKE or CMOKE in a given sample, thus cannot explain any of the three-fold and four-fold angular dependence in sample~Ni(111) and sample~Ni(001), respectively.

\begin{figure*}
\includegraphics{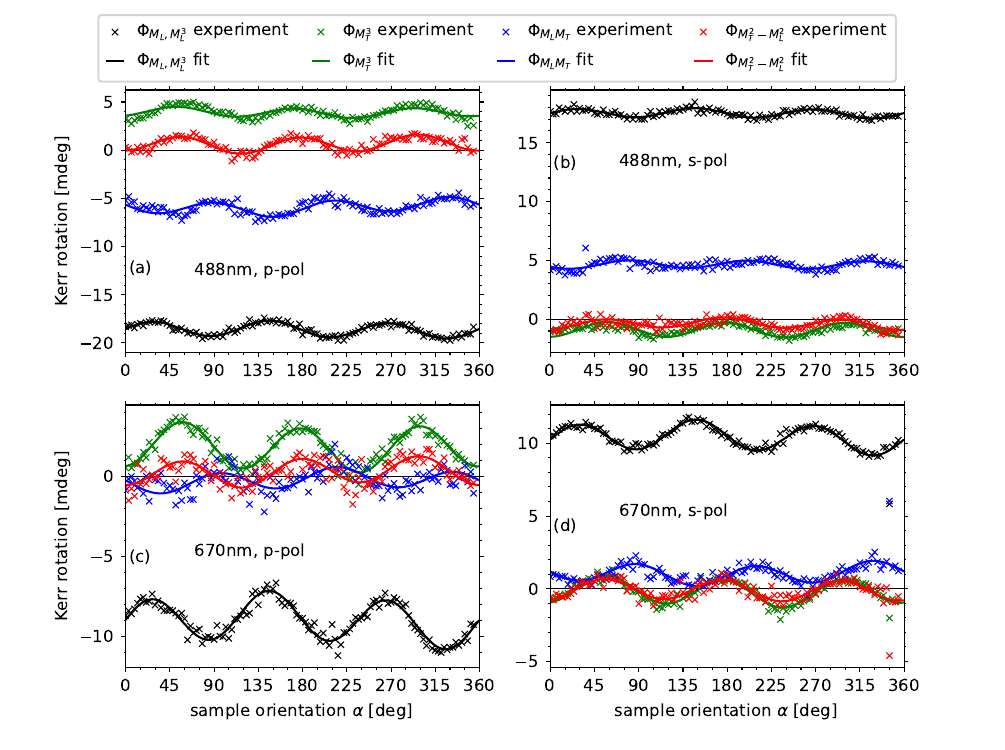}
\caption{Kerr rotation of eight-directional method measurement at an AoI of  45$^\circ$ for sample Ni(111) for (a) $p$-polarization and (b) $s$-polarization at a wavelength of 488\,nm and for (c) $p$-polarization and (d) $s$-polarization at a wavelength of 670\,nm. } 
\label{f:8dir_111RS_670_488}
\end{figure*}

\section{Eight-directional method measurements of Ni(111) sample with $s$- and $p$-polarization}
\label{app:Ni111_polarizations}

Table \ref{tab:sample_Ni111_contributions} of the main text suggests the sign invariance upon polarization change of three-fold angular dependence amplitude for (111)-oriented cubic crystal structures. In order to experimentally demonstrate this, we perform another eight-directional method measurement of sample Ni(111) with $p$-polarization and $s$-polarization at a wavelength of 488\,nm as shown in Figs. \ref{f:8dir_111RS_670_488} (a) and \ref{f:8dir_111RS_670_488}(b), respectively and at a wavelength of 670\,nm as shown in Fig. \ref{f:8dir_111RS_670_488}(c) and \ref{f:8dir_111RS_670_488}(d), respectively. As can be seen in the graphs, it is only the sign of the offset that is affected by the change of the polarization of the incident light exactly as the theory predicts. The experimental data are fitted by the model of Eq.~\eqref{eq:simple_model_111}.

\clearpage
\bibliography{qmokefe}

\end{document}